\documentclass[useAMS,usenatbib,a4paper, fleqn]{mnras}
\usepackage{ae,aecompl}
\title[M33 giant molecular clouds]
{Comparing the Properties of GMCs in M33 from Simulations and Observations 
}
\author[Dobbs]
{C. L. Dobbs\thanks{E-mail:
dobbs@astro.ex.ac.uk}$^{1}$, E. Rosolowsky$^{2}$, A. R. Pettitt$^{3}$, J. Braine$^{4}$, E. Corbelli$^{5}$, J. Sun$^{6}$ \\
$^1$ School of Physics and Astronomy, University of Exeter, Stocker Road, Exeter, EX4 4QL, UK \\
$^2$ Department of Physics, University of Alberta, Edmonton, AB, Canada \\
$^{3}$ Department of Physics, Faculty of Science, Hokkaido University, Sapporo 060-0810, Japan\\
$^{4}$ Laboratoire dÕAstrophysiquede Bordeaux, Univ. Bordeaux, CNRS , B18N, all\'ee Geoffroy Saint-Hilaire, 33615 Pessac, France \\
$^{5}$ INAF-Osservatorio Astrofisico di Arcetri, Largo E. Fermi 5, 50125 Firenze, Italy \\
$^{6}$ Department of Astronomy, The Ohio State University, 140 West 18th Avenue, Columbus, OH 43210, USA \\}

\usepackage{amssymb}
\usepackage{amsmath}
\usepackage[pdftex]{graphicx}
\usepackage{epsfig}
\usepackage{multirow}

\begin{document}
\label{firstpage}
\date{\today}

\pagerange{\pageref{firstpage}--\pageref{lastpage}} \pubyear{2012}

\maketitle

\begin{abstract}
We compare the properties of clouds in simulated M33 galaxies to those observed in the real M33. We apply a friends of friends algorithm and CPROPS to identify clouds, as well as a pixel by pixel analysis. We obtain very good agreement between the number of clouds, and maximum mass of clouds. Both are lower than occurs for a Milky Way-type galaxy and thus are a function of the surface density, size and galactic potential of M33. We reproduce the observed dependence of molecular cloud properties on radius in the simulations, and find this is due to the variation in gas surface density with radius. The cloud spectra also show good agreement between the simulations and observations, but the exact slope and shape of the spectra depends on the algorithm used to find clouds, and the range of cloud masses included when fitting the slope. Properties such as cloud angular momentum, velocity dispersions and virial relation are also in good agreement between the simulations and observations, but do not necessarily distinguish between simulations of M33 and other galaxy simulations. Our results are not strongly dependent on the level of feedback used here (10 and 20\%) although they suggest that 15\% feedback efficiency may be optimal. Overall our results suggest that the molecular cloud properties are primarily dependent on the gas and mass surface density, and less dependent on the localised physics such as the details of stellar feedback, or the numerical code used.
\end{abstract} 


\section{Introduction}
Molecular clouds are the sites of star formation in galaxies. As such, understanding the properties, formation and nature of molecular clouds in galaxies is central to determining star formation rates, cluster ages and age distributions, and the distribution of stars and star formation in galaxies. In the past, much work has been done to try to establish the properties of clouds in numerical simulations. However very little work has directly tried to reproduce a specific GMC population in a given galaxy. Here we attempt to do this by using Smoothed Particle Hydrodynamics (SPH) simulations of the M33 galaxy and CO (2-1) observations of GMCs in M33.

Until relatively recently, we were only able to observe molecular clouds in a very small sample of galaxies including the Milky Way, M33 and the LMC. However ALMA has enabled molecular clouds to be analysed in a much greater sample of galaxies, and surveys such as PAWS have achieved high resolution in nearby galaxies \citep{Schinnerer2013,Elmegreen2017,Tosaki2017,Faesi2018,Sun2018,Utomo2018,Kaneko2018}. Consequently we are now starting to build up a more complete picture of molecular clouds and their variation both between galaxies, and between different environments within the same galaxy. Observations have found small, but notable differences between GMCs in different galaxies. For example, \citet{Hughes2013} find that GMCs in M51 and the Milky Way are larger and have higher velocity dispersions compared to M33 and the LMC. Results from the PHANGS-ALMA survey \citep{Sun2018} find an increase in surface density, velocity dispersion and pressure for more massive galaxies. Observations also show differences between clouds in different environments, e.g. spiral arms, inter-arm regions and galaxy centres \citep{Colombo2014}. There are also some differences in star formation efficiencies \citep{Kennicutt2007,Bigiel2008,Kreckel2018}, and in particular M33 appears to have a particularly high star formation efficiency (for example compared to the Milky Way, \citealt{Gardan2007}).
Understanding what gives rise to these differences should help to explain what processes are important for star formation across different galaxies and environments.

Numerical simulations have also investigated the properties of GMCs in galaxies. Many have used models based approximately on the Milky Way. These simulations have produced results that are basically in general agreement. Simulations are able to produce realistic mass spectra, velocity dispersions, cloud rotations and virial parameters (e.g. \citealt{Tasker2009,Dobbs2011new,Dobbs2013,Khoperskov2016,Grisdale2018}). More recently, some studies have started to investigate the variation of GMC properties and star formation with galactic 
environment. \citet{Nguyen2018} investigate galaxies with different rotation curves, showing that typical cloud properties were not dependent on even quite extreme changes to the rotation, but the characteristics of the largest clouds / associations were effected. Adding a spiral potential (but otherwise keeping the stellar and gas mass the same) is also seen to produce slightly larger clouds, and a factor of 2 increase in star formation rates \citep{Dobbs2011new,Nguyen2018}. \citet{Pettitt2018} investigate the role of tidal interactions, comparing GMC properties at different stages of the interaction. Again the general properties of the clouds are not significantly affected by the interaction, but the tidally induced spiral arms do produce larger GMCs than occur when the galaxy is in isolation.

As well as modelling GMCs in galaxies, several works have also produced synthetic observations to directly compare with observed data \citep{Pan2015,Pan2016,Khoperskov2016,Duarte2016,Duarte2017}. Using radiative transfer modelling, \citet{Duarte2016} find that the CO traces only parts of greater underlying H$_2$ structures (see also \citealt{Smith2014}), and compare clouds in arm and inter-arm regions. \citet{Pan2015,Pan2016}  investigate the orientation of the galaxy, and differences in clouds as viewed in PPV and PPP space, finding that similar properties of clouds are traced by each, but the orientation can influence measurements of virial parameters of the clouds.  Both groups used algorithms from the observational community to extract the clouds, \citet{Duarte2017} used SCIMES \citep{Colombo2015}, and \citet{Pan2015} used CPROPS \citep{Rosolowsky2006}. In addition to cloud-extraction algorithms, observers have also studied galaxy properties simply with intensity weighted averages on a pixel by pixel basis along the line of sight \citep{Leroy2016}, which provides an additional way of comparing with simulations.

Although there are now many simulations investigating GMC properties, few have attempted to model actual galaxies. Doing so would provide a reference point for whether the simulations are accurately modelling galaxies and GMC formation and evolution. One exception is \citet{Renaud2015}, who model the Antennae system. However this is a fairly extreme case involving a galaxy collision leading to particularly massive clouds and clusters. \citet{Pettitt2018} compared the properties of GMCs formed in their simulations with M51, although the surface densities and interaction were not chosen to particularly match the M51 interaction. 

In this paper we compare properties of GMCs in simulations of M33 with observed GMCs in M33. M33 is a Local Group member which is smaller in size and mass than the Milky Way, and has a less clear spiral pattern compared to the Milky Way and M51. M33 is characterised by a relatively weak spiral structure, which exhibits multiple spiral arms rather than a grand design pattern.  We successfully reproduced the large scale spiral structure of M33 in previous work \citep{Dobbs2018}. Our models showed that the spiral structure can be reproduced by transient gravitational instabilities in the stars and gas, in agreement with observational results suggesting that M33 is undergoing a first approach with M31 \citep{Patel2017,vanderMarel2018}. We also found that quite strong levels of stellar feedback were required to best reproduce the spiral structure. Here we determine cloud properties from these previous simulations using two different algorithms, one of which CPROPS, is applied to both the simulations and observations. We also compare the properties of the simulated galaxies with M33 using the pixel by pixel analysis method of \citet{Sun2018}. As a nearby galaxy, M33 has been well studied at fairly high resolution in CO \citep{Engargiola2003,Gratier2010,Druard2014,Corbelli2017}, and as such we are able to use previous data and cloud catalogues for our comparisons.

\section{Method}
\subsection{Details of Simulations}
We compare GMC properties from three simulations designed to model M33, which were performed using the Smoothed Particle Hydrodynamics codes \textsc{sphNG} \citep{Bate1995} and \textsc{gasoline2} \citep{Wadsley2017}. The details of the simulations are described much more fully in \citet{Dobbs2018}, and they are also listed in Table~\ref{tab:simtable}.

All simulations were set up based on the observed properties of M33 from \citet{Corbelli2014}. 
We include one \textsc{sphNG} simulation (SPHNG20), which is based on the model labelled `Highres' in \citet{Dobbs2018}. The simulation is very similar to the model `Highres' with a couple of differences. In addition to the HI gas profile described in \citet{Dobbs2018}, we also included an exponential profile designed to mimic the molecular gas profile \citep{Corbelli2014}. We do not differentiate between atomic and molecular gas (since we cannot well resolve H$_2$ formation), we simply add an extra mass component to the disc. The radial profile of this extra component is 2.2 kpc \citep{Corbelli2014}. We decrease the mass of the rest of the disc by a factor of 10\%. This gives a central gas concentration which was not present in the simulations in \citet{Dobbs2018}. The added mass from the exponential profile is $2.8\times10^8$ M$_{\odot}$, similar to the observed molecular gas \citep{Corbelli2014}. The rotational velocities are recalculated according to the change in gas profile. Otherwise the physics included, including the cooling and heating, and the stellar feedback, are the same as \citet{Dobbs2018}. Stellar feedback is included using the simple prescription of \citet{Dobbs2011new}, whereby an amount of energy given by
\begin{equation}
E=\frac{\epsilon M 10^{51}}{160} ergs
\end{equation}
is inserted for each star formation event. Here $10^{51}$ ergs is the energy released by one supernova, $\epsilon$ is an efficiency parameter, and we assume that one massive star forms per 160 M$_{\odot}$ of stars formed.  A relatively high level of feedback is used with an efficiency of 20 \%. The simulation is set up with a dark matter halo, and stellar density profile the same as \citet{Dobbs2018}. The only other difference, aside from the exponential profile was that the value of the Toomre parameter, $Q$, was slightly lower, at around 0.93. Over time, after initial oscillations $Q$, tends to a value of around 1.2. This value is consistent across the disc until the gas surface density drops off (see \citealt{Dobbs2018} for discussion of the evolution of $Q$) . This model produces a slightly stronger spiral pattern which is in better agreement with the observed M33, and because of this, and the presence of more gas in the centre of the disc which matches observations, we mostly use this revised model for our cloud comparisons.

We also use two \textsc{gasoline2} simulations from \citet{Dobbs2018},  labelled GSLNfb10 and GSLNfb20 in that paper, and named GASOLINE10 and GASOLINE20 here. In \citet{Dobbs2018} we found that using either 10 or 20\% efficiency in the \textsc{gasoline2} simulations produced a reasonable match with observations (no equivalent run with 10\% feedback for \textsc{sphNG} was included). An efficiency of 10\% is typically used in \textsc{gasoline2} galaxy scale simulations. Unlike the \textsc{sphNG} calculation, feedback is treated as an entirely separate process to star formation, whereby star particles inject thermal energy into the surrounding ISM after their formation, rather than feedback originating directly from gas particles. The stellar feedback also has a separate efficiency to the star formation. This approach is common to standard sub-grid physics prescriptions in cosmological simulations. See \citet{Stinson2006} for details.

Both the \textsc{sphNG} and \textsc{gasoline2} simulations have similar resolution, the mass per particle is 409 and 440 M$_{\odot}$ in the two simulations respectively. This resolution allows us to consider clouds $\gtrsim 10^4$ M$_{\odot}$, which is similar to the masses to the observed range of clouds in \citet{Braine2018}. 

\begin{table}
\begin{tabular}{c|c|c|c}
 \hline 
Name of & Origin & Feedback & Mass per \\
simulation  & & efficiency & particle (M$_{\odot}$)  \\
 \hline
 SPHNG20 & Highres*+extra & 20 & 409 \\
  & gas component & & \\
GASOLINE10 & GSLNfb10* & 10 & 440 \\
GASOLINE20 & GSLNfb20* & 20 & 440 \\
\hline
\hline
\end{tabular}
\caption{Table showing the simulations used for the analysis presented here. *The simulation names in the second column are those are used in \citet{Dobbs2018}.}\label{tab:simtable}
\end{table}

\subsection{Details of Observations}
The CO(2-1) survey used to identify GMCs was carried out using the IRAM 30 m radio telescope, see \citet{Druard2014} for full details. The observed data covers the full galactic disc of M33 out to 7 kpc, and has a resolution of 12'', or 49 pc. Cloud catalogues are already published for the data \citep{Corbelli2017} and an analysis of cloud rotations is shown in \citet{Braine2018}. In Section~3.1 we take cloud properties from \citet{Corbelli2017} and \citet{Braine2018}. In Section~3.2 we have re-run CPROPS (see Section 2.3) on the CO(2-1) data to produce new plots which are shown in that section.

\subsection{Analysis of simulations}
We present analysis of the simulations at times of 419 Myr for the \textsc{sphNG} simulation, and 730 Myr for the \textsc{gasoline2} simulations. We choose these timeframes as they particularly closely resemble the observed structure of M33. As discussed in \citet{Dobbs2018}, the simulations periodically match the observations well at time intervals of $\sim150$ Myr when the orientation and shapes of the main arms agree with observations. However we also checked the cloud properties at earlier and later times (covering 140 Myr in the \textsc{sphNG} simulation, and 400 Myr in the \textsc{gasoline2} simulation) and did not find any significant variation in the cloud properties.

We use two different methods to analyse GMC properties. The first is the friends of friends cloud-finding algorithm used in \citet{Dobbs2015} and more recently by \citet{Pettitt2018}. This algorithm works by first selecting gas particles whose SPH density exceeds a given density threshold, $\rho_{crit}$. From this list of denser particles, we then selects particle which lie within a given distance ($l_{crit}$) of any other particles. There is some degeneracy between these two parameters, as choosing a high $\rho_{crit}$ and low $l_{crit}$ can give similar results to choosing a low $\rho_{crit}$  and high $l_{crit}$. Here we choose parameters to give a reasonable match with the observed cloud distribution. We also require that each cloud contains a minimum of 100 particles. This ensures each cloud is well resolved, and by chance approximately matches the completeness limit of the observations.  The algorithm is 3D in nature.
\begin{table*}
\begin{tabular}{c|cc|c|c|c}
 \hline 
Name of & Cloud & Relevant & Number & Maximum &  $\gamma$ (slope of mass \\
cloud sample  & selection  method & parameters & of clouds & cloud mass M$_{\odot}$ & function)  \\
 \hline
FFsphNGA & FoF & $\rho_{crit}=8$ cm$^{-3}$, $l_{crit}=15$ pc & 517 & 2$\times10^6$ & -2.27\\
FFsphNGB & FoF & $\rho_{crit}=5$ cm$^{-3}$, $l_{crit}=20$ pc & 727 & 1$\times10^7$ & -1.81 \\
CPsphNG20 & CPROPS &  $\rho_{crit}=5$ cm$^{-3}$ & 867 & 1$\times10^6$ &  -$1.93$ \\
FFGASOL10 & FoF & $\rho_{crit}=8$ cm$^{-3}$, $l_{crit}=15$ pc & 444 & 3.5$\times10^6$ &-1.95\\
CPGASOL10 & CPROPS &  $\rho_{crit}=5$ cm$^{-3}$ & 434 & 1$\times10^6$ & -$1.70$ \\ 
CPGASOL20 & CPROPS &  $\rho_{crit}=5$ cm$^{-3}$ & 593 & 1.5$\times10^6$ & -$1.66$ \\ 
Observations & CPROPS &  $\rho>5$ cm$^{-3}$ & 564 & 2$\times10^6$ & -1.65 (Braine et al. 2018) \\
 &  &  & 485 & 3$\times10^6$  & -$1.59$ (see Section~3.2)\\
\hline
\end{tabular}
\caption{Table showing the cloud populations found from the \textsc{sphNG} simulation, the \textsc{gasoline2} simulations and the observations. Those which include `sphNG' are from the SPHNG20 calculation, GASOL10 from the GASOLINE10 simulation, and GASOL20 from the GASOLINE20 simulation. Clouds are selected using a Friends of Friends algorithm (FoF) and CPROPS. For the CPROPS results, the name of the simulation tends to be used (since there is only one CPROPS result for each simulation) rather than cloud sample. The final column indicates the slope of the mass spectra for the simulations and observations. The uncertainties on $\gamma$ are $\pm0.2$. For the cloud mass spectra, the results using the FoF algorithm are compared using a fitting approach which gives a very similar match to \citet{Braine2018}. All the CPROPS results are shown in Section 2.3 where the CPROPS algorithm was applied identically to both the simulations and observational data. This produced slightly different results to \citet{Braine2018}, though well within the uncertainties given by CPROPS.}\label{tab:cloudtable}
\end{table*}

We also use the cloud-finding algorithm CPROPS \citep{Rosolowsky2006}, which was applied to both the observational data and the simulations. 
For the simulations, we create mock observational data by projecting the simulations into celestial coordinates using the orientation parameters given in \citet{Koch2018}.  For each particle, we assume it represents molecular gas if it has a hydrogen density larger than $5~\mathrm{cm}^{-3}$ (so we are using CPROPS in the optically thin mode).  If so, we convert its mass to an equivalent amount of CO(2-1) emission using a CO-to-H$_2$ conversion factor of $X_\mathrm{CO}=4\times 10^{20}\mbox{ cm}^{-2}/(\mbox{K km s}^{-1})$ \citep{Gratier2017} and a line ratio of CO(2-1)/CO(1-0)=0.8 \citep{Druard2014}.  The precise value of the conversion factor does not affect the derived molecular cloud properties since we assume the same value for scaling the emission back to estimated mass. However, it does affect the recovery of low mass or low surface brightness molecular emission since these will be found below the noise levels of the observations. We then generate a mock data set matching the properties of the IRAM CO(2-1) map.  Specifically, we map the CO emission from the SPH particles onto the coordinate grid using Gaussian kernels.  The spatial scale of the kernel is set by the SPH smoothing length and the spectral width is set by the particle temperature. We then convolve this map by mock instrumental response represented by a Gaussian beam with a width of $11ÕÕ$ and a boxcar channel width of $2.6~\mathrm{km~s}^{-1}$ matching the IRAM CO(2-1) data. The instrumental response is significantly broader than the smoothing kernels applied directly to the simulation data, so the precise kernel used in gridding the particle data does not affect the final results. Finally, for each data set, we create a mock noise field matching the properties (noise level, spatial distribution) estimated from the signal-free part of the real data cube. These mock data sets then mimic the real observations, but the assumed beam lacks the sidelobe structure of the real observations.  Since we are focused on the properties of the compact CO emission peaks, the proper treatment of the instrumental response will not be critical to this study.

For each mock data set, we generate a GMC catalogue using the CPROPS algorithm \citep{Rosolowsky2006}. To identify emission, we include elements in the data cube that are larger than $5\sigma_{\mathrm{rms}}$ where $\sigma_{\mathrm{rms}}$ is the local noise level.  We then expand this mask into all connected elements in the data cube that are larger than $2\sigma_{\mathrm{rms}}$ in two consecutive channels. We reject regions that are smaller than 20 total pixels.  We search this masked emission region for local maxima that are $<3\sigma_{\mathrm{rms}}$ below the saddle point that connects them to a brighter local maximum.  The remaining local maxima define the GMCs in the catalogue, and we use a seeded watershed algorithm to assign the emission in the mask to the associated local maximum.  With this assignment, we calculate cloud properties as per \citet{Rosolowsky2006}, using the cloud properties extrapolated to the 0 K emission level.

\section{Results}

\begin{figure*}
\centerline{\includegraphics[scale=0.58, bb=350 200 500 520]{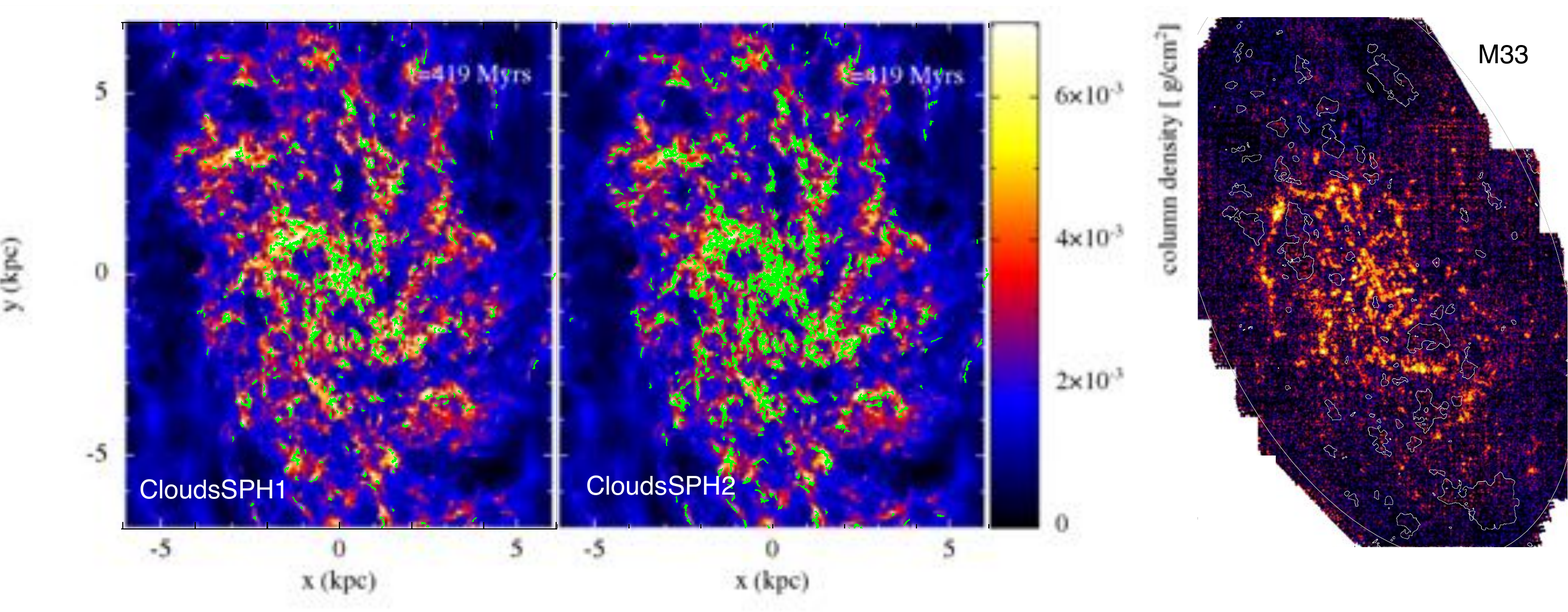}}
\caption{High density clouds (green) are overplotted on the total gas density for the SPHNG20 simulation in the left and middle panels. In our clumpfinding algorithm we adopt a critical density of $\rho_{crit}=8$ cm$^{-3}$  and length of  $l_{crit}=15$ pc in the left panel, and $\rho_{crit}=5$ cm$^{-3}$ and $l_{crit}=20$ pc in the middle panel. The right panel shows a CO (2-1) map of M33 from \citet{Druard2014}.} 
\label{fig:cloudmap}
\end{figure*}

\begin{figure}
\centerline{\includegraphics[scale=0.33, bb=100 120 500 650]{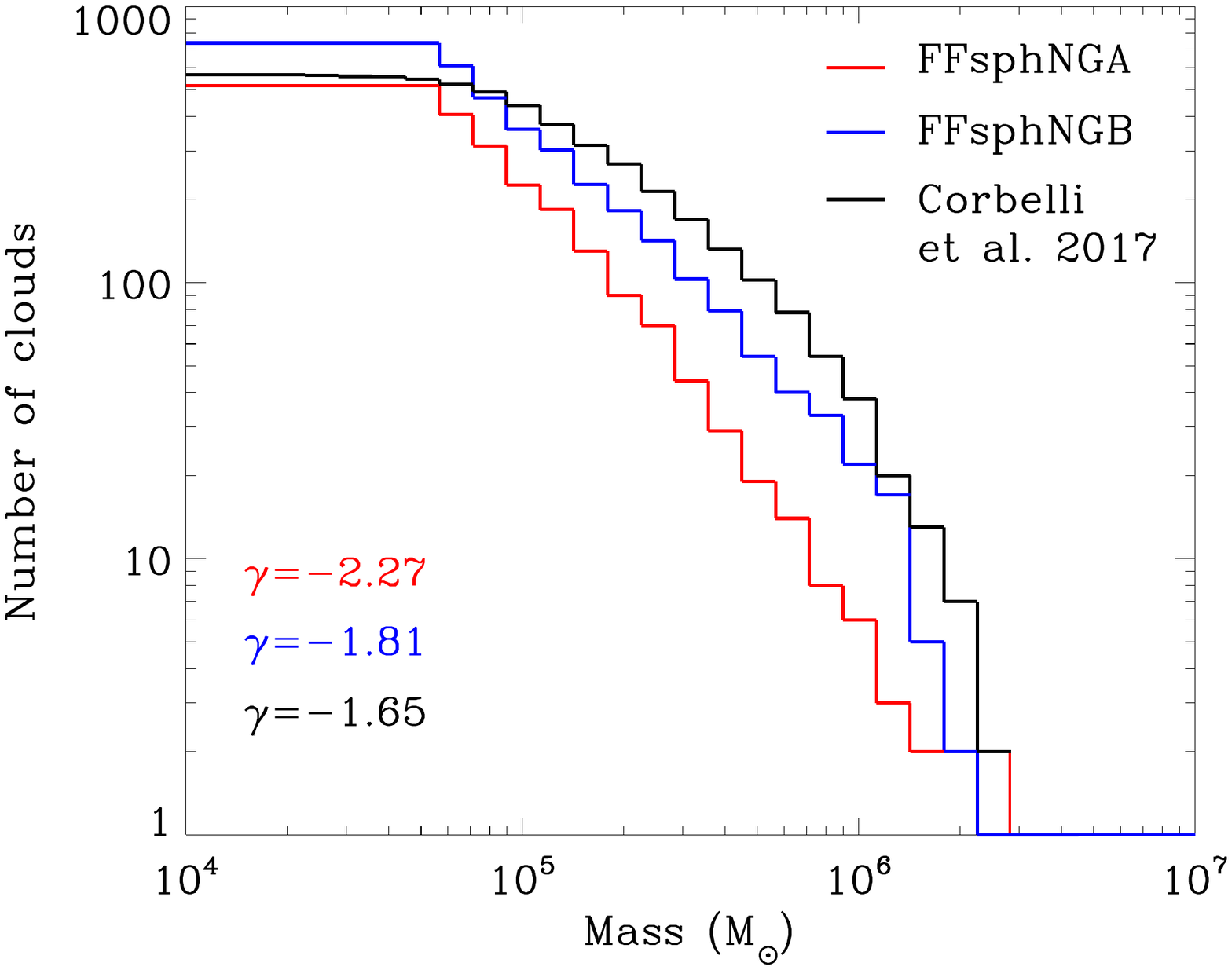}}
\centerline{\includegraphics[scale=0.33, bb=100 150 500 600]{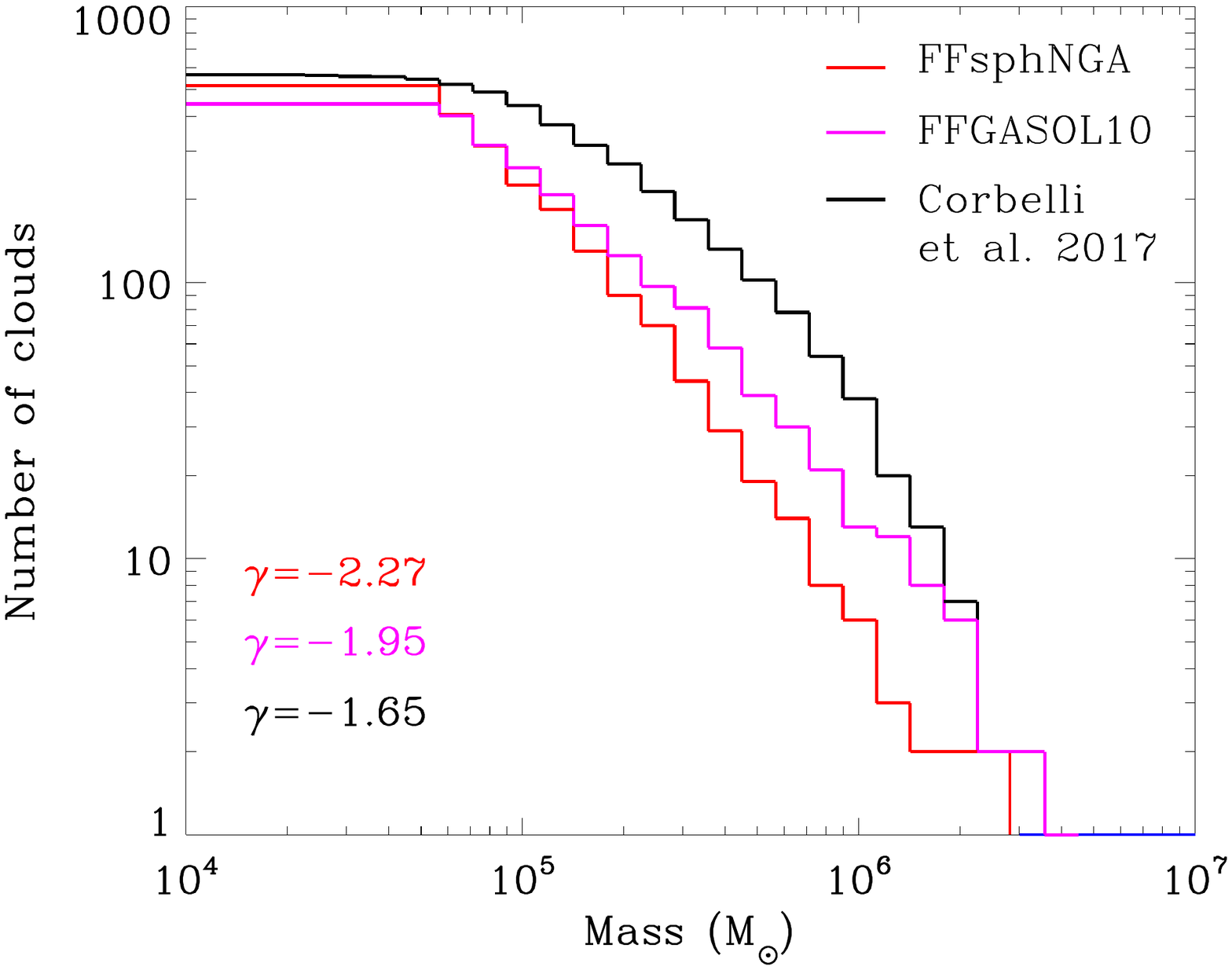}}
\caption{In the top panel, we show cloud mass spectra from the SPHNG20 calculation using different parameters for the friends of friends algorithm. In the lower panel we compare clouds found in the SPHNG20 and GASOLINE10 simulations, using the friends of friends algorithm with the same parameters. The black line in each panel shows the observed cloud mass spectra found using CPROPS.}
\label{fig:masssp}
\end{figure}

\begin{figure}
\centerline{\includegraphics[scale=0.34]{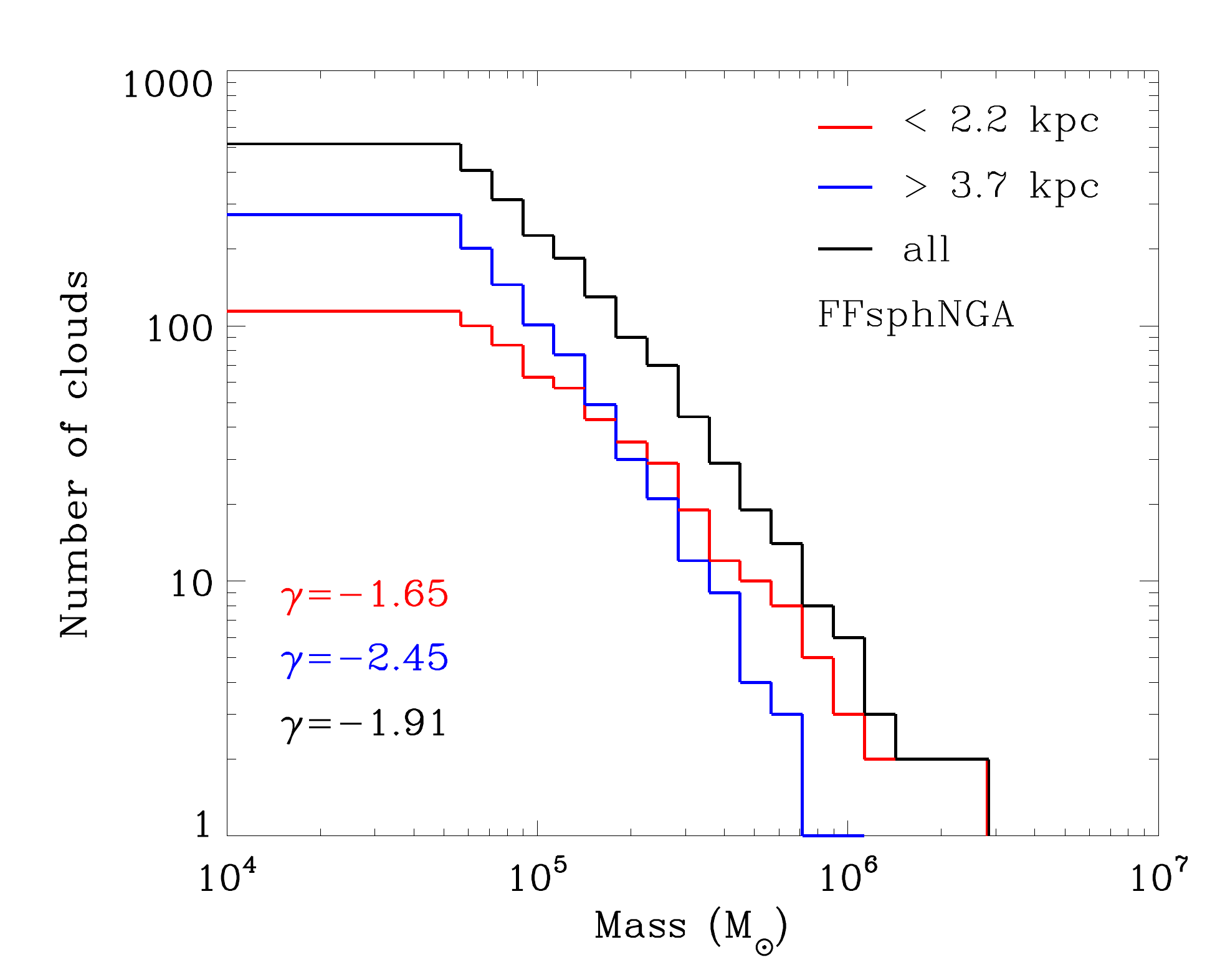}}
\centerline{\includegraphics[scale=0.34]{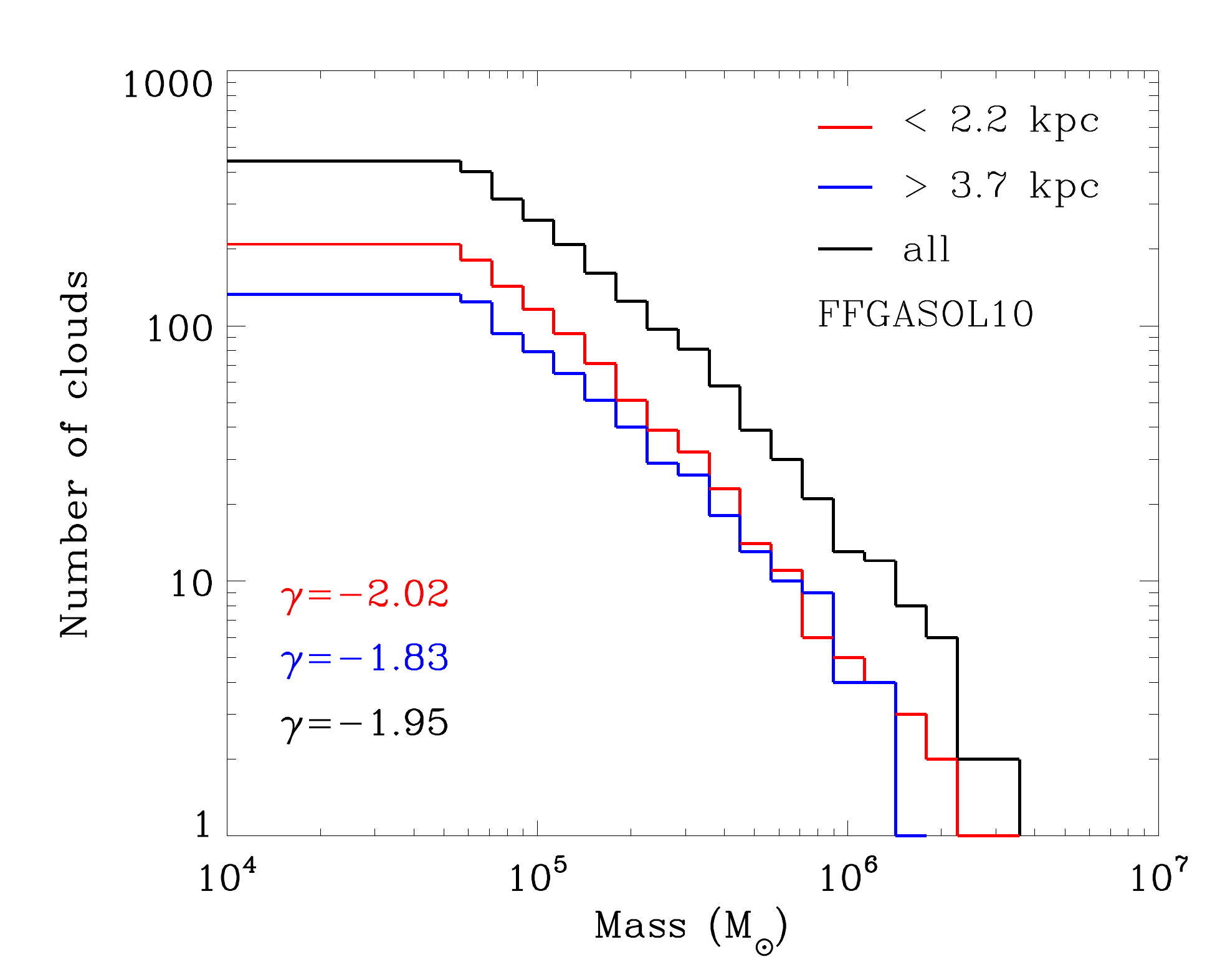}}
\centerline{\includegraphics[scale=0.34]{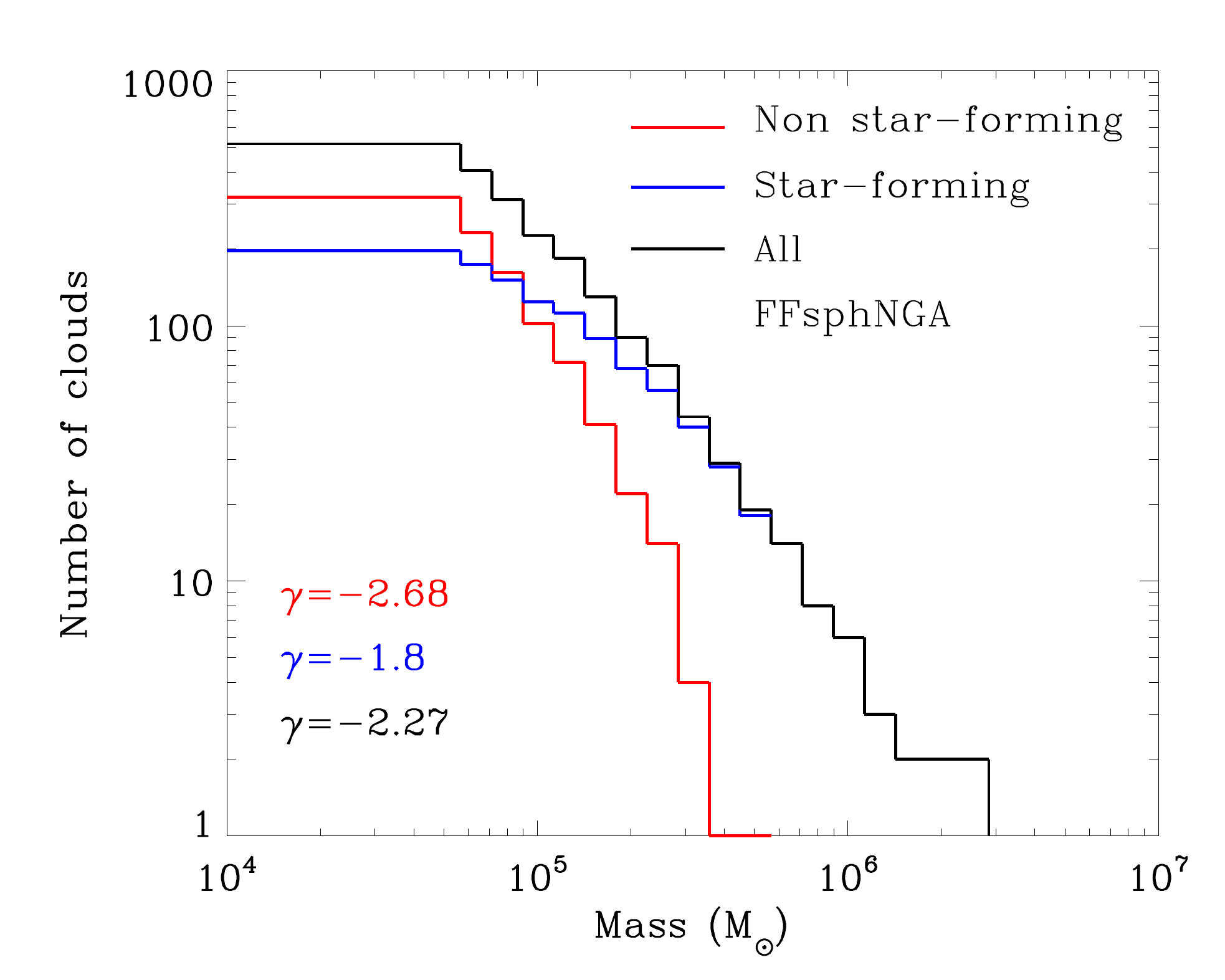}}
\caption{Cloud mass spectra are shown for clouds divided according to radius ($<$ 2.2 and $>3.7$ kpc) for the FFsphNGA (top) and FFGASOL10 (middle). The lower radius clouds display a shallower slope and a higher maximum cloud mass compared to the high galactic radius clouds, which is in agreement with observations \citep{Braine2018}. This is less evident for the GASOLINE10 clouds where the initial galaxy setup did not include an additional gas component at the centre representative of molecular gas. The lower panel shows spectra for the clouds divided according to whether they have recently formed stars (see text for details) for the SPHNG20 clouds.}
\label{fig:sprad}
\end{figure}

\subsection{GMCs determined using friends of friends algorithm}
We first consider the populations of clouds identified from the simulations using the friends of friends algorithm. We apply this algorithm to the SPHNG20 and GASOLINE10 models (with 20 and 10\% feedback efficiency). We choose the 10\% efficiency \textsc{gasoline2} model because the cloud properties were a slightly better match to the observations than those from GASOLINE20. However the differences between the \textsc{gasoline2} models were not that large. We discuss different feedback efficiency models in Section 3.1.3, and we also include both \textsc{gasoline2} simulations in our comparisons with CPROPS in Section 3.2.

In all our results we use the total density to determine the clouds, as molecular gas evolution cannot generally be resolved in galactic scale simulations \citep{Duarte2015}.
We show the parameters used to find the clouds, and highlight some overall properties of the resulting cloud populations in Table~\ref{tab:cloudtable}. 
We explored results using two different sets of parameters for the friends of friends algorithm, applied to the \textsc{sphNG} simulation. We first take $\rho_{crit}=8$ cm$^{-3}$ and $l_{crit}=15$ pc, and secondly use $\rho_{crit}=5$ cm$^{-3}$ and $l_{crit}=20$ pc, and call the two resulting populations of clouds `FFsphNGA' and `FFsphNGB' respectively. In both cases the densities of the clouds are quite low, but this largely reflects the relatively low gas densities in the disc, the resolution of the simulation, and the injection of feedback at high densities. The clouds extracted from the catalogue of \citet{Corbelli2017} tend to have densities which are mostly above $\rho_{crit}$. For the first set of parameters, we found 517 GMCs, with a maximum cloud mass of $2 \times 10^6$ M$_{\odot}$ whilst for the second set of parameters, we found 727 GMCs, and one massive outlier cloud with a mass of $10^7$ M$_{\odot}$, which was located at the centre of the galaxy. 

We show the total gas column density with the clouds selected according to the two sets of parameters in Figure~\ref{fig:cloudmap}, for the SPHNG20 simulation. We also show the M33 CO (2-1) map from \citet{Druard2014}. The simulated galaxy has been rotated according to the inclination and position angles observed for M33 from \citet{DeVauc1991}. A comparison of the large scale spiral structure shows reasonable agreement between the simulations and observations. There is a prominent arm to the left of the galaxy, although it is not quite as extended in the simulation compared to the actual M33. There is also a long spiral arm feature extending to the top region of the galaxy, and some short spiral features in the lower part of the plot. In both the simulated galaxy and M33, the clouds appear more preferentially located in the spiral arms, and at the centre of the galaxy.

\subsubsection{Mass spectra}
In this section we show the mass spectra from the simulations and observations. We also fit a truncated power law to the mass function as described in \citet{Pettitt2018}.  In this formalism the slope of the mass function is denoted $\gamma$, and we list $\gamma$ for the mass distributions found using the friends of friends algorithm, and the observations in Table~\ref{tab:cloudtable} (as well as denoting them on the relevant figures). The results for the observations in this section use the cloud catalogue of \citet{Corbelli2017}, and for fitting the spectra, we used clouds with masses $>6.1 \times 10^4$ M$_{\odot}$, which is the same cut off as used by \citet{Braine2018}. 

In Figure~\ref{fig:masssp} we show the cloud mass spectra for the two sets of cloud parameters, the SPHNG20 and GASOLINE10 simulations, and the M33 clouds (the same figure for M33 alone is shown in \citealt{Braine2018}).  The top panel compares the spectra from the SPHNG20 simulation using the two different sets of parameters for the clump-finding algorithm.  We see that although there is some broad similarity between the observed and simulated cloud spectra, there are some differences. Both the total number of clouds, and the maximum cloud mass agree almost exactly between the observation and simulations (see also Table~\ref{tab:cloudtable})\footnote{note that although there are clouds at larger radii than seen in the observations, even if discounting these the total number of clouds and maximum cloud mass from FFsphNGA matches the observations best compared to the other simulated cloud populations.}. However the mass spectra for the simulated clouds have a different shape to the observations, the latter appearing curved whereas the simulations give a cloud population with a clear power law slope. This also means that the total mass of clouds is less than observations; we could better match the total mass by changing the criteria for cloud selection criteria, but this would then produce clouds which are too massive compared to those observed. FFsphNGA is also clearly steeper than the observed spectrum. FFsphNGB matches the middle part of the spectrum better but produces too many clouds, and one very massive ($10^7$ M$_{\odot}$) cloud not present in the observations. 

In the lower panel of Figure~\ref{fig:masssp}, we compare the spectra for the \textsc{sphNG} and \textsc{gasoline2} simulations (FFsphNGA and FFGASOL10). The spectrum for FFGASOL10 is flatter than that of FFsphNGA. The reason could be the way stellar feedback is inserted. In the SPHNG20 simulation, stellar feedback is inserted  when gas becomes dense, and therefore affects lower mass clouds in exactly the same way as large mass clouds. If feedback is delayed, as is the case for the \textsc{gasoline2} simulations, more clouds may be able to reach higher masses, hence the spectra at high masses matches the observations better. The downside of the FFGASOL10 population is that the total number of clouds is too low compared to the actual M33, and there is a dearth of low mass clouds. Changing the friends of friends parameters did not help as the maximum cloud mass increased but the total number of clouds only changed a small amount. Similarly to the spectra from SPHNG20, the \textsc{gasoline2} simulation GASOLINE10 is unable to reproduce the curved shape seen in the observed mass spectrum. 

The fitted mass functions to the spectra all show steeper slopes compared to the observations. However this is largely a consequence of the mass threshold used to fit the spectra, which we chose to be the same as \citet{Braine2018}, and the curved shape given by the observations. If we take a larger mass limit, then steeper slopes are found for all the data, but the observations then lie within the range of the simulated cloud spectra. The observational results produced with CPROPS also have uncertainties, and when we show further analysis with CPROPS in Section 2.3, slightly higher slopes in better agreement with the simulations are obtained.

For the remainder of this section we only consider the FFsphNGA population, since the further results we show are applicable for both the FFsphNGA and FFsphNGB populations. Although FFsphNGB produces a better agreement with the observed spectrum, the clouds in FFsphNGA (as indicated in Figure~\ref{fig:cloudmap}) tend to be more compact. The clouds in both populations tend to be less dense than those observed but this discrepancy is worse for the clouds from FFsphNGB. The number of clouds in FFsphNGA is also in better agreement with the observations. Similarly changing the parameters for the clumpfinding algorithm did not produce significantly better results for the \textsc{gasoline2} simulation so we also use only one realisation of the clumpfinding algorithm for this simulation (FFGASOL10).

To further compare with \citet{Braine2018} we divided clouds according to where they are located in the disc. \citet{Braine2018} find that the spectrum becomes steeper for clouds at larger radii, and massive clouds are only found at small radii (their Figure~7). We show the mass spectra for FFsphNGA divided according to the same radial bins as \citet{Braine2018} in Figure~\ref{fig:sprad} (top). We observe a very similar trend to that seen in the observations, i.e. a steeper spectrum at large radii, and a shallower spectra at small radii. The spectra also extends to higher masses for the lower radii clouds similar to the observations. In Figure~\ref{fig:sprad} (middle panel) we also show the spectra for the clouds found in GASOLINE10, FFGASOL10, for the different radial bins. The slopes for the different radial bins are similar to that of the total population of clouds (in fact the slope is slightly steeper for the clouds at low galactic radii). Thus there is no indication that more massive clouds are present only at lower radii, which is dissimilar to the observations. This difference is likely due to the absence of the extra gas in the centre of the galaxy, which is present in the SPHNG20 but not GASOINE10. We further checked the cloud mass spectra for the `Highres' calculation, which used \textsc{sphNG} but similarly did not have an extra gas component at the centre, and similar to GASOLINE10, this showed little difference in the spectra for clouds in different radial bins. In the `Highres' calculation there is still a slight tendency for massive clouds to be nearer the centre, which may be due to the increased stellar surface density towards the centre. \citet{Corbelli2019} also suppose that more massive clouds can be formed at the centre due to the fast rotation of the disc with respect to the
spiral arm pattern that allows extra growth of the clouds as they cross the arms. We have not tested
this explicitly in the simulations.
 
We then divided the clouds into `star-forming' and `non star-forming'. Star forming clouds were selected as all those which contained any gas which was at least 1000 K. Non star-forming clouds contained only gas below 1000 K, whereas feedback in star-forming clouds heats localised regions to temperatures greater than 1000 K. The feedback algorithm injects energy according to the amount of star formation and so this increases the temperature locally within a cloud. The dense regions observed in CO are of course much denser (thus colder) than the regions that can be followed in a simulation.
The density of gas in the clouds ($\gtrsim10$ cm$^{-3}$) is such that in the absence of star formation or stellar feedback, all gas should be below 1000 K (e.g. \citealt{Field1965,Dobbs2008}), therefore any gas which exhibits temperatures higher than this will have been heated by star formation activity and so similar to the observations, these clouds show observable signs of recent star formation. We show the spectra of these two groups of clouds in Figure~\ref{fig:sprad} (lower panel) which can again be compared with \citet{Braine2018}. Note that \citet{Braine2018} had an additional class of embedded star formation, but we do not include this class as it would be too difficult to extract from the simulations. Similar to the observations, the non star-forming clouds show a steeper spectrum, and the star-forming clouds a shallower spectrum compared to the spectrum for the total number of clouds. In both the simulations and observations, the difference in the spectra for star-forming and non star-forming clouds is greater than the radial variation. There are a couple of reasons why the more massive clouds preferably contain star formation. The more massive clouds are statistically more likely to contain dense regions, and thus star formation. Also, the clouds likely gain further mass as they start forming stars \citep{Dobbs2013}, so non-star forming clouds may simply be at an earlier stage in their lifetime when they are lower masses.

Although not shown, we also looked at the location of the star-forming and non star-forming clouds, and found that the non star-forming clouds tend to be at larger galactic radii. This indicates that more massive clouds tend to occur more towards the centre of the galaxy, and tend to be more likely to exhibit star formation compared to their lower mass counterparts. This again is likely related to the increased gas surface density in the disc, although other factors may be relevant such as stronger stellar spiral arms and a stronger galactic potential towards the centre of the galaxy, which lead to more readily star forming clouds. As non-star forming clouds at both small and large radii are found to go on to produce star formation (see Section 3.1.4), the finding of more star-forming clouds towards the centre suggests that the timescales for clouds to form, and for stars to form within them, is shorter compared to the outer regions of the galaxy. We looked at the evolution of the clouds prior to and after their selection at the time of 419 Myr. Gas which went on to form clouds towards the centre of the galaxy tended to be denser, indicating that clouds are able to form quicker in the inner parts of the disc.

\subsubsection{Cloud sizes and velocity dispersions}
In this section we show cloud masses, sizes and velocity dispersions. We show here figures for the FFsphNGA population, but also show results from the FFGASOL10 population in the appendix. In Figure~\ref{fig:massrad} we show the radius versus mass of the clouds from the simulation and observations. The size of the simulated clouds tends to be larger than the observed ones, and only tend to match the observed clouds at low masses. The simulated clouds tend to exhibit constant densities, whereas for the observed clouds, the higher mass clouds tend to be higher overall density. As shown in the appendix, the distribution is similar for FFGASOL10, and thus not strongly dependent on the feedback prescription. The densities of the clouds will instead depend on the parameters for the cloud-finding algorithm, and likely the resolution of the simulations. In reality, a massive cloud may have gas which exhibits a range of densities, whereas in the simulation there is a limited range of densities within a cloud, so clouds tend to be clustered around the minimum densities required for the clouds to be selected by the cloud-finding algorithm.

We also plot the variation of cloud mass, and the velocity dispersions of the clouds versus galactic radius in Figure~\ref{fig:massvel}, again for both the observed and simulated clouds. Overall the range of cloud mass and velocity dispersions are comparable between the observations and simulations, although again there are some differences. The simulated clouds extend to larger galactic radii than the observed clouds, likely because in the real M33 there is a sharper drop off in surface density at radii $>8$ kpc, although in the simulation the large majority of clouds still lie at radii $<8$ kpc (for FFGASOL10 which has a sharper drop off in surface density the clouds match better the observed radial distribution). The simulated clouds also include clouds with larger velocity dispersions compared to the observations. 

Both the observed and simulated clouds show some variations with galactic radius, although these variations are weaker for the observed clouds. There is a tendency for massive clouds to occur nearer the centre of the galaxy in both the real and simulated galaxy (see also Figure~6 of \citealt{Braine2018}), again likely a consequence of the higher gas surface density towards the centre. The velocity dispersion of the simulated clouds also tends to on average decrease with galactic radius (see again Figures~5 and 6 of \citealt{Braine2018}). This trend reflects that the more massive clouds tend to have higher velocity dispersions, and the more massive clouds are situated towards the centre of the disc, but we note that stellar feedback also varies across the disc and this also determines the velocity dispersion. The velocity dispersion of the FFGASOL10 clouds are in better agreement with the observations, if anything they underestimate the maximum observed velocity dispersions. The FFGASOL10 clouds also show a slight decrease in the velocity dispersion with radius. The FFGASOL10 masses do not show any particular dependence on radius, which likely again reflects the flat density profile of the gas.

\begin{figure}
\centerline{\includegraphics[scale=0.4]{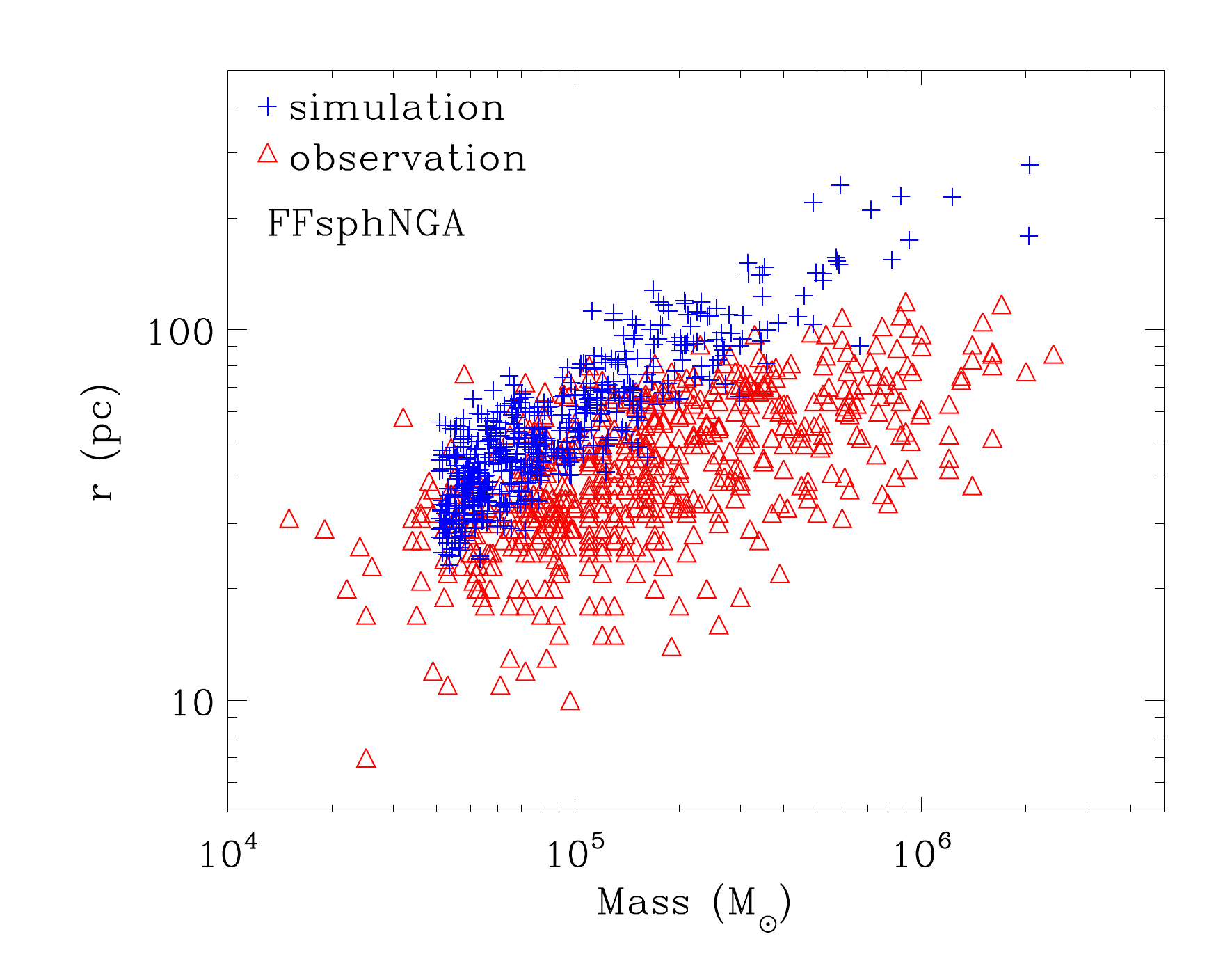}}
\caption{The radius of the clouds are plotted against their mass for the FFsphNGA population and the observations. The simulated clouds tend to show more constant densities compared to the observed clouds, and the more massive clouds tend to be too extended compared to the observed clouds.}
\label{fig:massrad}
\end{figure}

\begin{figure}
\centerline{\includegraphics[scale=0.4]{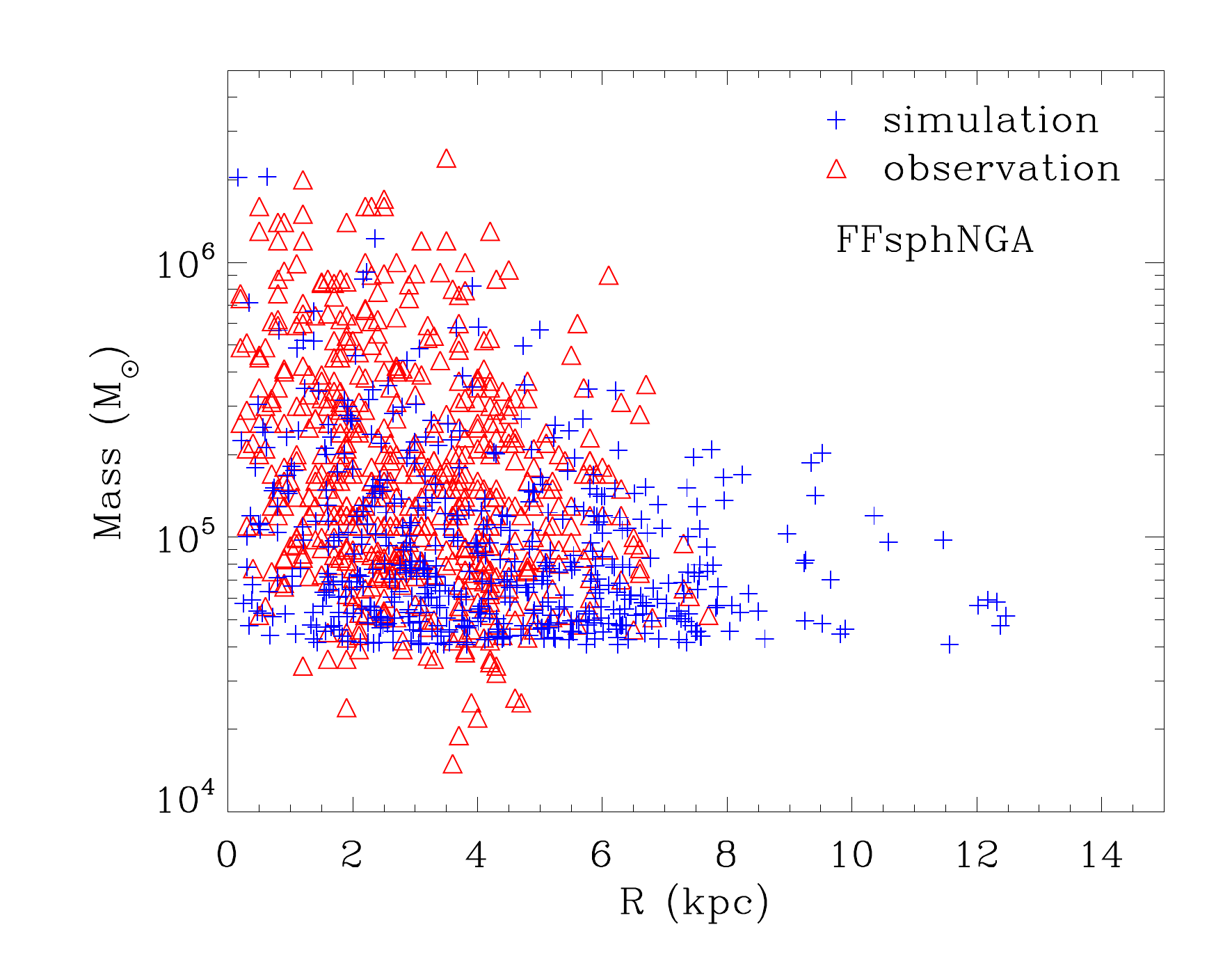}}
\centerline{\includegraphics[scale=0.4]{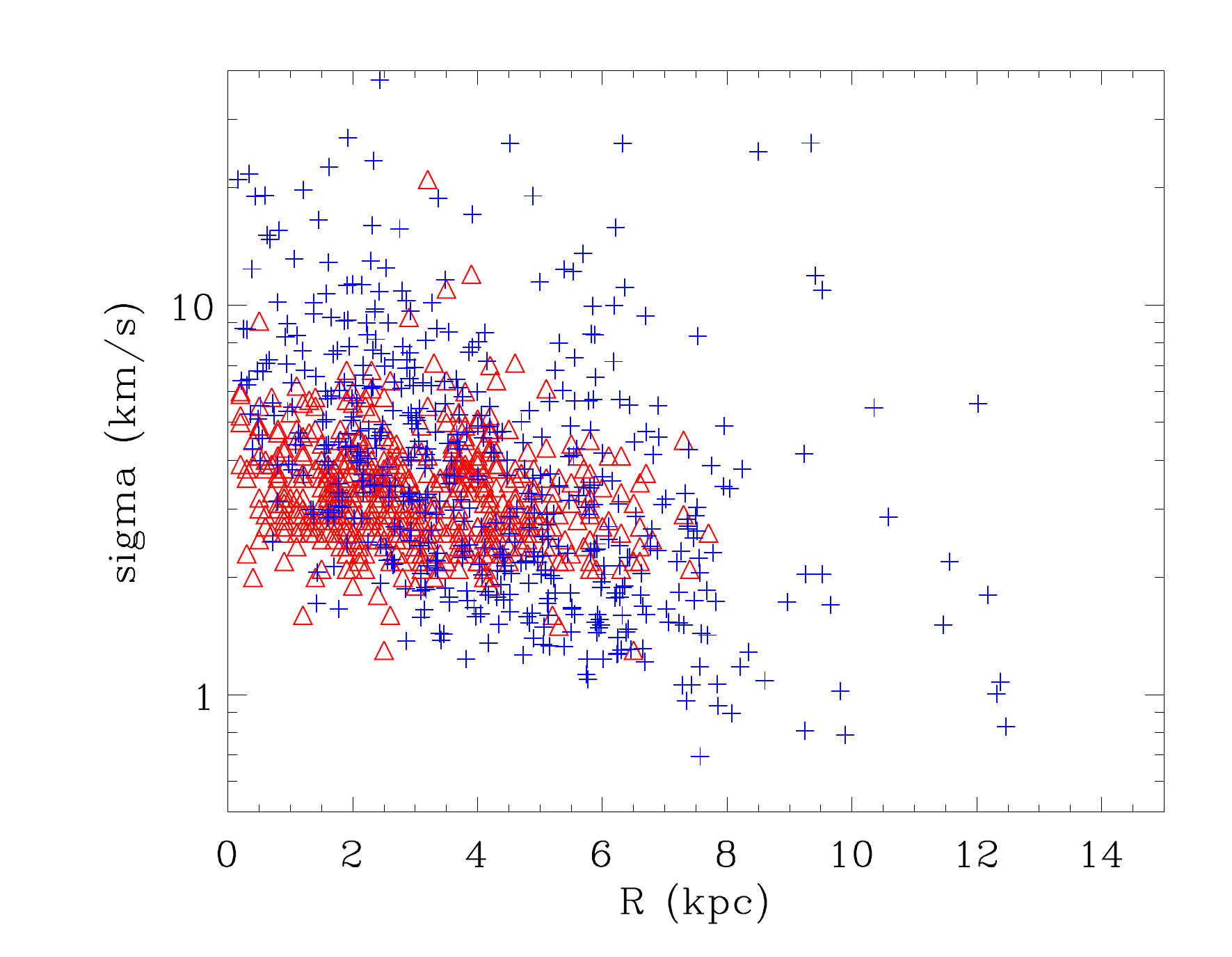}}
\caption{The masses (top) and velocity dispersions (lower) of the clouds are plotted versus galactic radius for the population FFsphNGA. Both the simulated and observed clouds show a decrease in the cloud mass and velocity dispersion versus galactic radius although this is more pronounced for the simulations. The clouds also exhibit fairly similar values, although there are more simulated clouds with high velocity dispersions.}
\label{fig:massvel}
\end{figure}

\begin{figure}
\centerline{\includegraphics[scale=0.4]{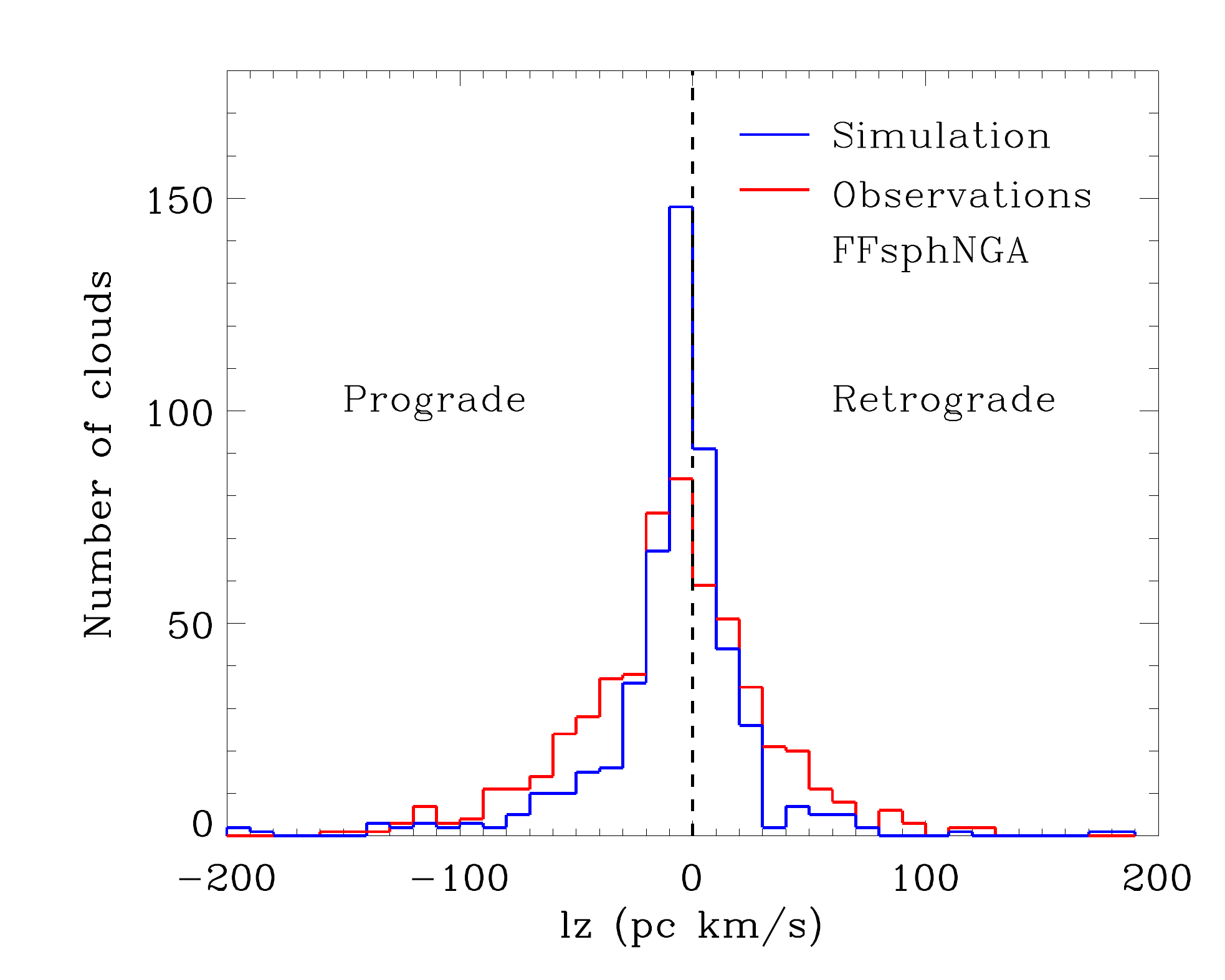}}
\caption{The distribution of angular momenta is shown for FFsphNGA and the observations. The distributions are fairly similar although the simulated clouds have a higher peak at low (prograde) angular momentum.}
\label{fig:rotation}
\end{figure}

\subsubsection{Cloud rotation}
Cloud rotation has increasingly been used as a diagnostic of the physics of molecular cloud formation and evolution. \citet{Dobbs2008} and \citet{Tasker2009} highlighted the substantial fraction of retrograde clouds seen in both observations and simulations and suggested there is an indication that cloud-cloud collisions are important in producing retrograde clouds. In contrast when the cloud population consists of clouds which are strongly self-gravitating, the fraction of retrograde clouds is  very low and less than observed \citep{Dobbs2008,Dobbs2011new}. The galactic potential and spiral arms may also influence the rotation (\citealt{Mestel1966}, Braine et al. submitted). We show the distribution of cloud angular momenta for the simulated and observed clouds in Figure~\ref{fig:rotation}. For the simulated clouds we calculate the intrinsic angular momentum of the clouds, whereas for the observations, the angular momentum is based on the velocity gradients across the clouds. Both the simulated and observed clouds exhibit a very similar range of angular momenta, with the large majority of clouds exhibiting angular momenta of $l_z<100$ pc km s$^{-1}$. The peak of the distribution is also in the same location for both the observed and simulated clouds, indicating that the preference for both the simulated and observed clouds is to have a small net amount of angular momentum which corresponds to prograde rotation (albeit that for these low values this probably does not resemble actual rotational motion). The peak is higher for the simulations, and the distribution narrower compared to the observations. However the higher angular momenta values shown in \citet{Braine2018} are the least reliable, and these are associated with clouds  with lower signal to noise ratios.
The distribution of angular momentum is very similar for the \textsc{gasoline2} clouds (FFGASOL10, see appendix). We also plotted the angular momentum against cloud mass (not shown) and found that both the simulated and observed clouds displayed a similar trend of a gradual increase of angular momentum with increasing cloud mass (see e.g. \citealt{Braine2018} and \citealt{Dobbs2013}). 

We also compared the fraction of prograde and retrograde clouds in both the simulation and observations. For the SPHNG20 simulation, 35\% of the clouds exhibited retrograde rotation versus 39\% seen in observations. \citet{Dobbs2011new} found that retrograde fractions start to decrease as clouds become more gravitationally dominated, so the similar fractions between the simulation and observations suggest that self gravity has a similar role or impact in the simulated galaxy compared to the actual M33.  The fraction of retrograde clouds in the GASOLINE10 simulation is 43\%.

\subsubsection{Cloud lifecycle}
\citet{Corbelli2017} identify clouds at different stages of their lifecycle and use the number of clouds at each stage to estimate the cloud lifetime (following the procedure of \citealt{Gratier2012}). They identify 3 different types of clouds, clouds without star formation, clouds with embedded star formation, and clouds with young stellar clusters. One assumption of their model is that clouds which are identified as non-star forming will go on to form stars in the near future. Numerical simulations provide a unique opportunity to test this assumption, as it is possible to trace the evolution of the clouds. We simply divide the clouds form the FFsphNGA population into those that have clearly exhibited recent star formation, and those which display no evidence of star formation. We use the temperature threshold as an indication of recent star formation activity, the same as Section 3.1.1. We then follow the non-star forming clouds for a period of 12 Myr. Over this period we determine the maximum density at each timeframe for each cloud. If clouds reach the density for star formation to assume to occur, then we can say that these clouds did indeed go on to form stars. Otherwise the clouds may just be transitory objects which would produce no or minimal star formation. In practice, clouds that go on to form stars have a continually increasing maximum density indicative of gravitational collapse, whilst those that do not form stars have quite different behaviour, so it is relatively straightforward to distinguish between the possible evolutionary scenarios for the clouds. 

Of the clouds which are denoted as non-star forming, we find 86 \% of these clouds do go on to exhibit star formation. 
Of the remaining clouds, 1/3 show an increasing maximum density, likely indicating that they will form stars on a timescale greater than 12 Myr. The remaining 2/3 show a decreasing, or steady maximum density, giving no indication that they will form stars. There are a couple of caveats to our results. The clouds in the simulations are typically less dense than the observed clouds which could underestimate the number that go on to form stars. On the other hand, we do not include magnetic fields, which may delay or prevent star formation in clouds. Overall though, the simulations indicate that the majority of clouds would be expected to go on to form stars in a short time, and the proposed lifecycle of clouds presented in \citet{Corbelli2017} and other similar work is valid.

\subsubsection{Comparison with other simulations}
We also checked the cloud properties for the GASOLINE20 simulation, with 20\% feedback and listed in Table~\ref{tab:cloudtable}. This simulation was also found to have reasonable agreement with the large scale structure of M33. The clouds found in GASOLINE20 were also found to have reasonable agreement with the observed clouds. The main drawback of the GASOLINE20 population was that there were fewer clouds, presumably as feedback acts to break up the clouds more and the smaller clouds do not match our criteria for selecting clumps. 

We also compare our simulated cloud catalogues with previous simulations which did not model M33 to see how sensitive the cloud properties are to the M33 galactic setup. We compared the number of clouds to the simulations in \citet{Dobbs2013}, who modelled a Milky-Way like galaxy, and \citet{Pettitt2018}, who modelled a tidally interacting galaxy. In all cases, the simulations extend to similar radii. The number of clouds in these previous papers is significantly higher ($\gtrsim1000$) than found in our models of M33. 
To properly compare, we used exactly the same cloud-finding algorithm with the same parameters as in those papers. We still found that the number of clouds was around half that of the previous work, with similar resolution calculations. When using the same cloud-finding parameters, the maximum cloud mass in our simulated M33 is also a factor of 10 or so lower compared to the other simulations.
Thus we conclude that it is the specific setup of the M33 model, including the surface density of the gas, and the gravity from the stars and dark matter, that produces a smaller number of lower mass clouds.  
Other properties, such as the rotation of the clouds, and their velocity dispersion, are similar compared with previous work, suggesting that these are not particularly dependent on the specific M33 setup, but rather the physics which is present in the simulations (particularly stellar feedback and cooling and heating).

\subsection{GMCs selected using the CPROPS algorithm}
\begin{figure*}
\centerline{\includegraphics[scale=0.7]{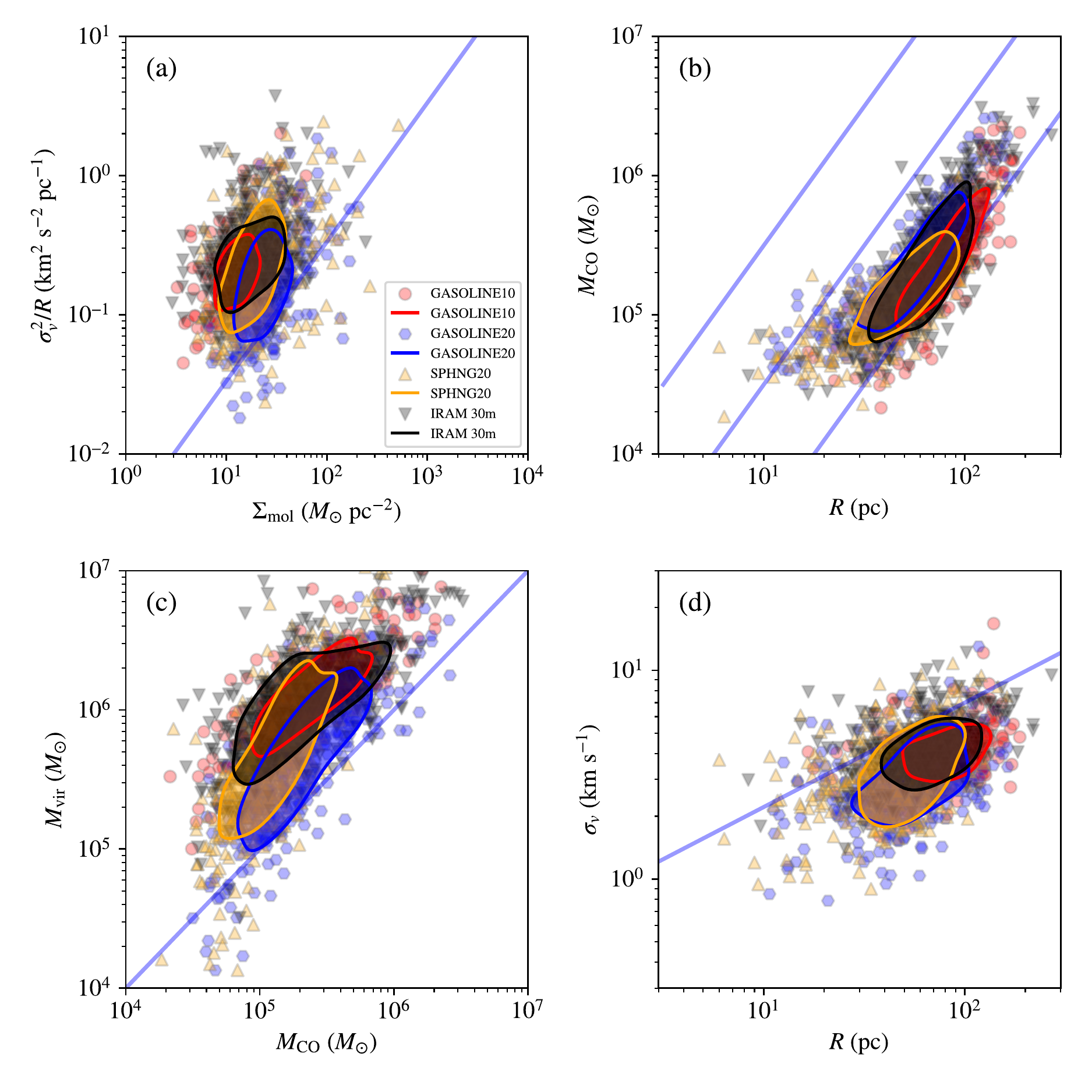}}
\caption{Cloud properties are shown for cloud found using the CPROPS algorithm. The panels show the virial relation (top left), mass (top right), virial mass (lower left) and velocity dispersion (lower right). The lines show a virial parameter of 1 (top left), a linear relation (lower left), constant surface densities of 10, 100 and 1000 M$_{\odot}$ pc$^{-2}$ (top right) and $\sigma_v =1.10 R^{0.38}$ (lower right) \citep{Larson1981}.}
\label{fig:cprops}
\end{figure*}

\begin{figure}
\centerline{\includegraphics[scale=0.7]{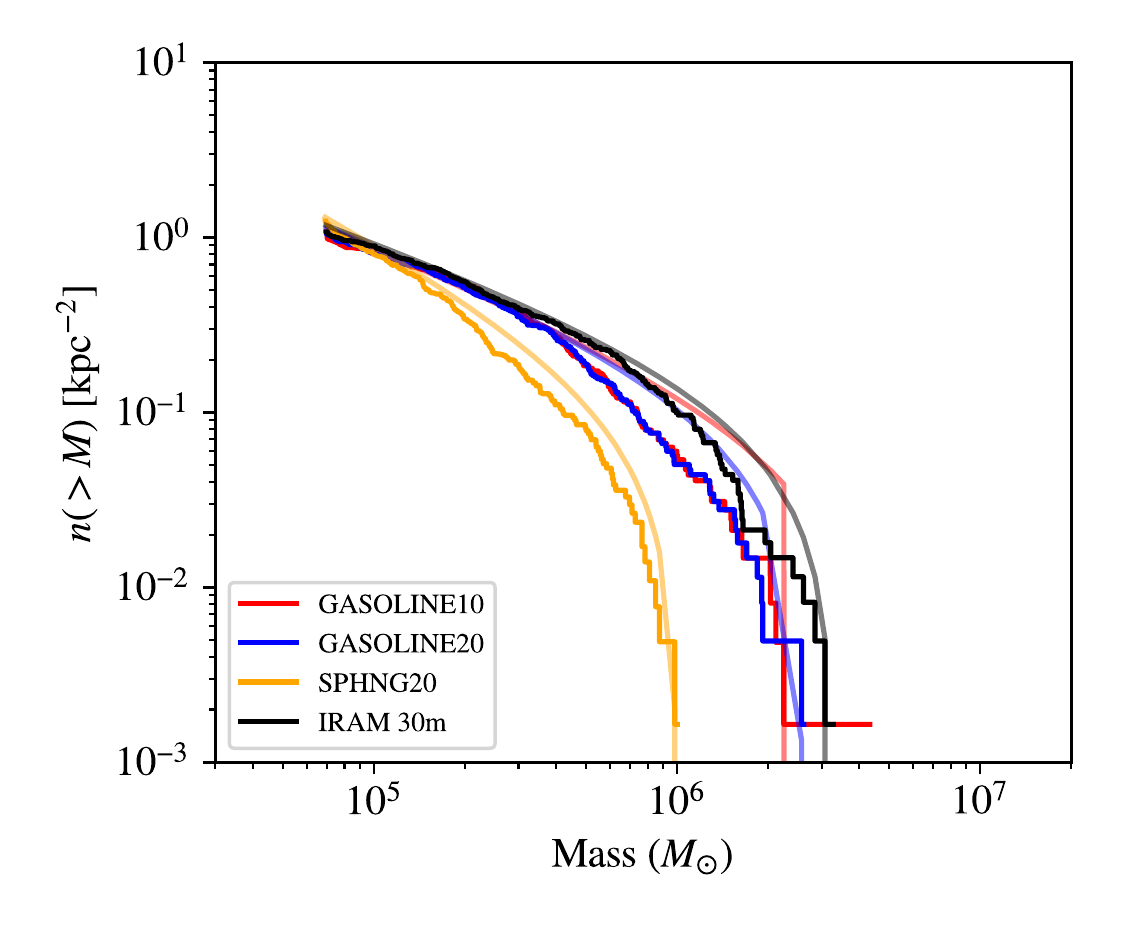}}
\caption{Mass spectra are shown for the simulations and observations, where the clouds are found using the CPROPS algorithm.}
\label{fig:cprops_sp}
\end{figure}

\begin{figure*}
\centerline{\includegraphics[scale=0.7]{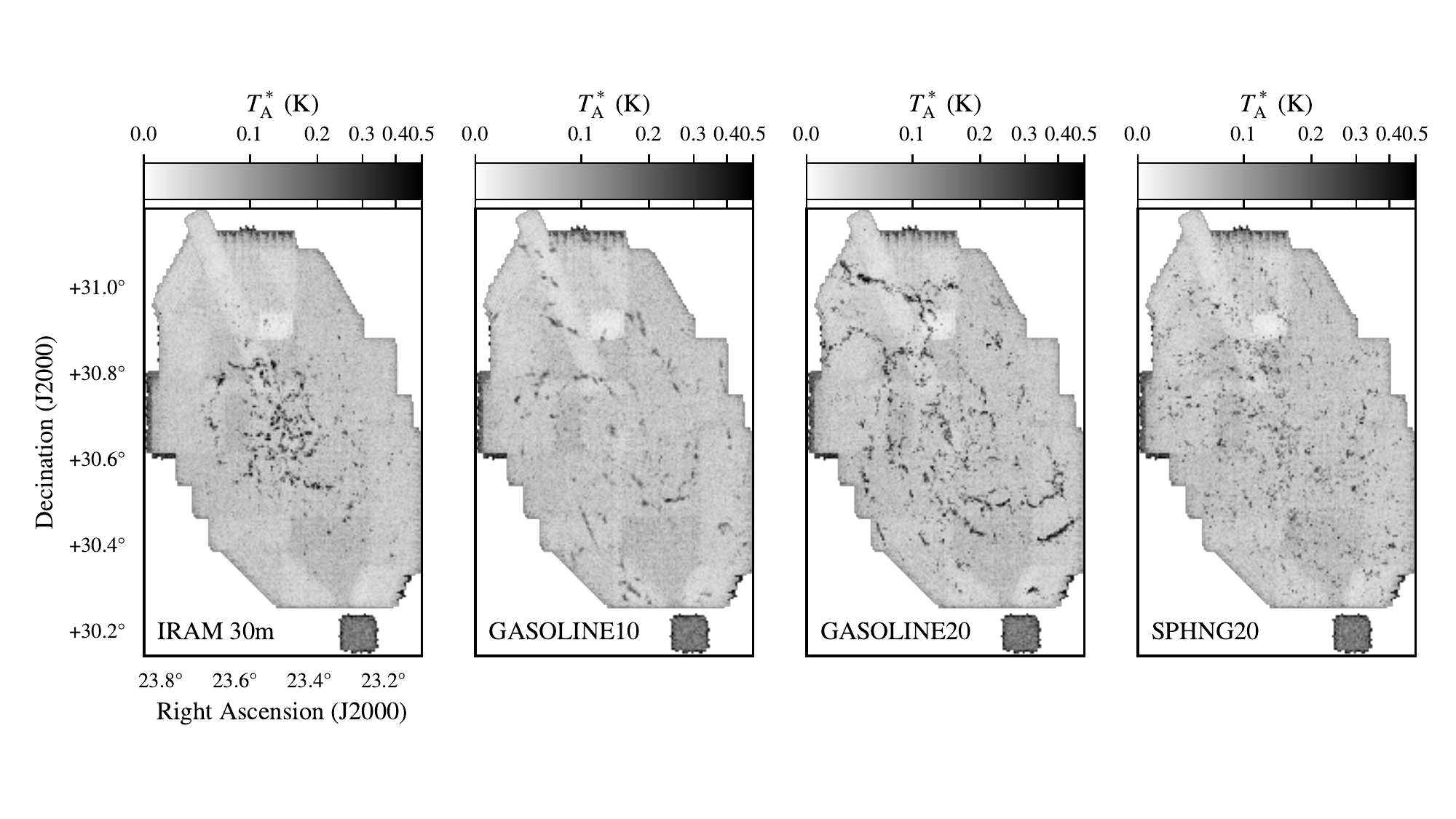}}
\caption{The left hand panel shows the IRAM 30 CO map of M33, then CO maps are shown for the simulated M33 galaxies in the other panels.}
\label{fig:cpropsmap}
\end{figure*}

\begin{figure}
\centerline{\includegraphics[scale=0.4]{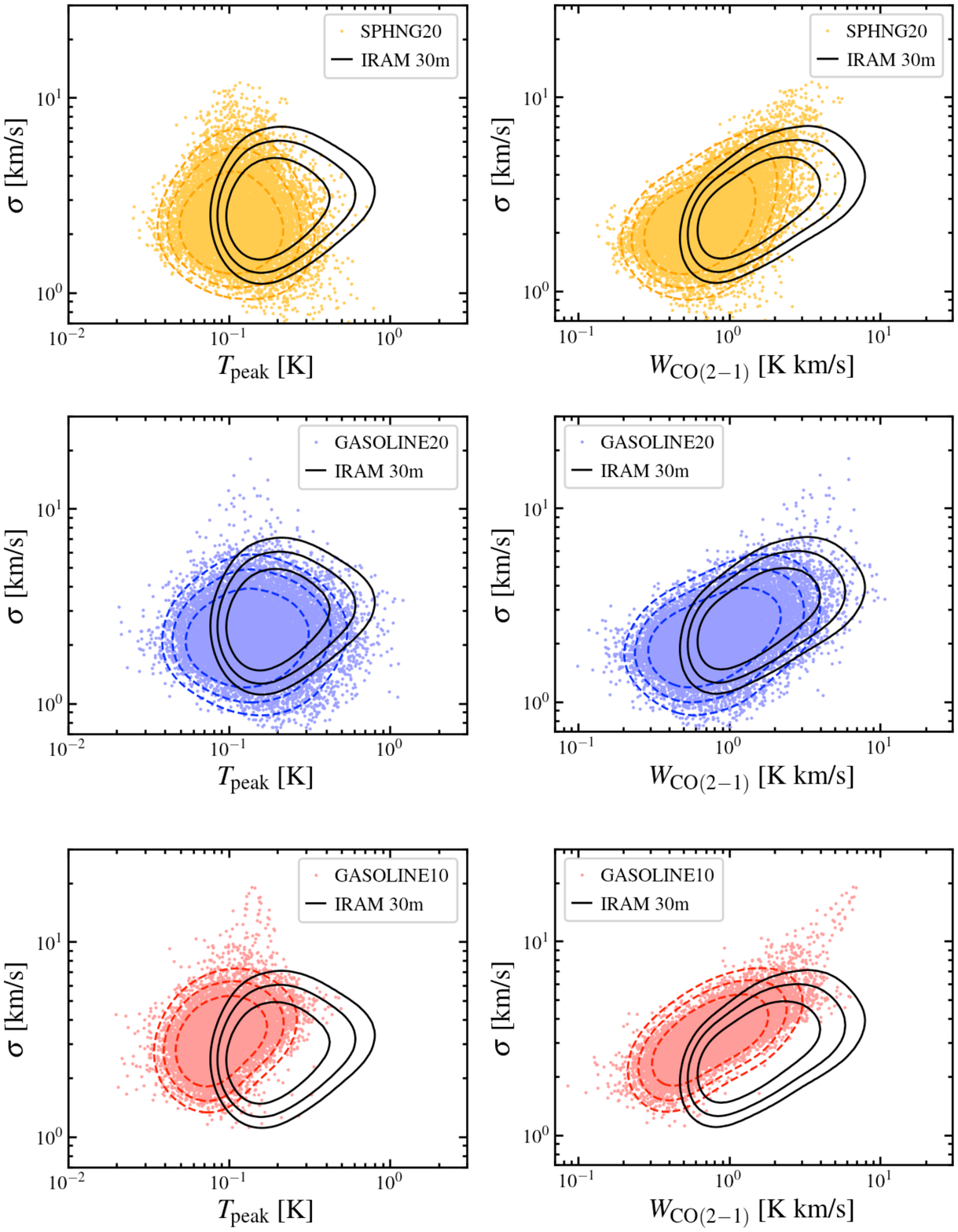}}
\caption{Correlations between pixel-by-pixel measured molecular gas properties are shown for the three simulations and observations. The left panels show velocity dispersion against peak temperature, and the right panels velocity dispersion against integrated emission. The simulations occupy a similar region of the parameter space to the observations, particularly the models with 20\% efficiency. }
\label{fig:pixel}
\end{figure}

In this section, we compare properties of clouds found in the simulations and observations using CPROPS. This analysis represents a truer comparison of the simulated and observed data compared to the previous section, as the same algorithm is used for each. 
We present cloud properties found using the CPROPS algorithm in Figure~\ref{fig:cprops}. The results are shown for the SPHNG20, GASOLINE10 and GASOLINE20 simulations, the cloud populations listed as  CPsphNG20, CPGASOL10 and CPGASOL20 respectively in Table~\ref{tab:cloudtable}. In particular we show the virial parameter versus surface density (top left), cloud mass versus radius (top right), virial mass (lower left) and velocity dispersion (lower right). There is considerable overlap between the clouds found in the simulations and those found in the observations for all the properties. In most cases the distribution of the cloud properties for the \textsc{sphNG} and \textsc{gasoline2} simulations with 20\% feedback are very similar. The distribution of cloud properties for the GASOLINE10 simulation tends to be shifted in comparison.
The range of the observational clouds, as indicated by the contours, appears to lie between the simulations with 10 and 20 \% feedback. This is consistent with our findings in \citet{Dobbs2018}, where we did not find a clear preference for 10 or 20 \% feedback from the simulations we ran. It also indicates that the differences between the simulations and observations are less than the differences which occur for a relatively small change in the feedback (a factor of 2 in the level of feedback). Figure~\ref{fig:cprops} shows that the simulated clouds with lower feedback tend to be more massive and larger sizes compared with higher feedback. 

For the virial relation (top left panel, Figure~\ref{fig:cprops}), the simulations and observations occupy a similar parameter space, with virial parameters a little above 1. Again this may reflect the clouds having similar density and the choices for the cloud-finding algorithm.  The clouds in the simulations and observations tend to have similar surface densities (top panels), like the friends of friends algorithm this is a consequence of the cloud-finding algorithm and the density thresholds used. Observations of other galaxies also tend to find a similar distribution of clouds when plotted on the virial relation.

In Figure~\ref{fig:cprops_sp} we show the cloud mass spectra for the clouds found using CPROPS. We fit the mass distributions using the method of \citet{Freeman2017}, who fit mass distributions using a power law distribution with an exponential truncation.  Figure \ref{fig:cprops_sp} shows the complementary cumulative mass distribution functions for for the four different catalogues and the respective best-fitting mass distribution.  We fit the entire distribution above $3\times 10^4~M_\odot$ based on the smallest mass clouds recovered through the observed area.  However, we have not done a full completeness test on these data for recovery as the analysis is intended to highlight a differential comparison between the mock data sets and observations.
Again, the simulated and observed clouds show reasonable agreement, and there are not large differences between the cloud properties. The number of clouds found is extremely similar for the observations and \textsc{gasoline2} simulations with 20\% feedback. The maximum cloud mass in each of the simulations lies within a factor of 2-3 of the maximum cloud mass of the observations. The spectrum for SPHNG20 is quite steep compared with the observations (seen also with the friends of friends algorithm). The stellar feedback is quite effective in this simulation, which may limit the formation of more massive clouds (although helping to particularly well reproduce the global spiral pattern). Figure \ref{fig:cprops_sp} also shows that there is more scatter in the simulations compared to observations. This could be a consequence of limited resolution in the simulations, e.g. not resolving star formation which might increase the velocity dispersion, inserting feedback in overly large regions of gas so that the velocity dispersions seen in the clouds are too high, not resolving low mass clouds (Figure~4). There may also be limitations with the observations, such as failing to resolve lower density, or particularly dense regions. As shown in the next section though, at least some of the scatter comes from the CPROPS algorithm, as when we compare the simulations and observations on a pixel by pixel basis, there is only slightly more scatter compared to the observations. 

It is also evident that the shape of the cloud mass spectrum from the simulated galaxies is much closer to the observed clouds in Figure~\ref{fig:cprops_sp}, where CPROPS is applied. This shows that the shape of the mass spectrum, whether it is a single or multiple power law, or a more log-normal shape, is attributable to the nature of the clump-finding algorithm.  As CPROPS is a clustering algorithm, it will tend to group smaller objects together, leading to fewer low mass clouds and more high mass clouds compared to a linear slope. 

In Figure~\ref{fig:cpropsmap} we show mock CO emission maps of the simulations, and the CO map for M33. No one of the simulations show a particularly good agreement with the M33 map, the main difference being that the emission extends to larger radii in the mock emission maps compared to the actual M33. It's not completely clear why this difference occurs given that the comparisons of the total density maps \citep{Dobbs2018} are similar and the clouds extracted using the friends of friends algorithm (Figures~\ref{fig:massvel} and A1) do not extend much further in the simulations compared to the observations. For the actual M33, the conversion of HI to molecular gas appears particularly inefficient outside the central 3 or 4 kpc \citep{Gardan2007}, so when using the total density we may overestimate the emission at large radii. Figure~\ref{fig:cloudmap} also suggests that there is gas in the simulations at larger radii but it is not necessarily identified as GMCs.

\subsection{Comparing the simulations and observations using a non cloud-decomposition approach}
So far our analysis has relied on using a technique to separate the CO emission, or total gas density in the simulations, into distinct entities or clouds. An alternative approach is to use a pixel by pixel analysis \citep{Leroy2016, Sun2018}, which is particularly suitable for data at marginal resolution and has the advantage that there is no dependence on how the CO emission is allocated to the identified GMCs. In this approach, cloud properties are calculated over a size scale according to the beam size (typically 10's pcs in surveys of nearby galaxies), matching the typical size of GMCs. 
We apply this pixel by pixel approach (for the full details, see \citealt{Sun2018}) to the data cubes which were made for the CPROPS analysis, for both the simulations and observations. 

We show plots of the velocity dispersion against CO (2-1) line peak temperature, $T_\mathrm{peak}$, and line-integrated intensity, $W_\mathrm{CO(2-1)}$, in Figure~\ref{fig:pixel}. Similar to the CPROPS analysis, we find that the simulated data and observations occupy a similar region of the parameter space.  In particular the points from the simulations overlap substantially with the observational data in these parameter spaces. We also see similar trends to the CPROPS results. The sphNG20 and GASOLINE20 simulations, both with a higher level of feedback produce a very similar distribution of points, but the GASOLINE10 points are offset from the other simulations. The pixel by pixel analysis suggests that the higher feedback simulations better reproduce the observed molecular gas properties, as shown in Figure~\ref{fig:pixel}.

\section{Discussion}
We have compared the properties of GMCs formed in simulations designed to reproduce M33, with GMCs in the actual M33. We use two methods to identify GMCs in the simulations, a friends of friends algorithm, and CPROPS, a commonly used method of observers. We overall find good agreement between the clouds in the observations and the simulations. In particular the total number of clouds, and the maximum cloud mass are in very good agreement (around at most a factor of two) with the observations using both methods. The slopes of the mass spectra are also in broad agreement. We also manage to reproduce similar differences between the mass spectra for clouds at small versus large radii, and with and without clear star formation, compared to the observations. We attribute this radial dependence at least partly to the molecular gas component at low radii in M33, which was also included in the SPHNG20 simulation which produced similar trends. The overall gravitational potential is also generally higher in the inner part of the galaxy, due to higher stellar as well as gas surface densities, which may be reflected in the cloud properties.

We find a noticeable difference in the mass spectra determined using the friends of friends algorithm and CPROPS. The friends of friends algorithm tends to produce a simple power law mass spectrum, whereas CPROPS produces a more curve shape to the power spectrum, or a multiple power law. Differences between algorithms have been noted before. \citet{Khoperskov2016} compared the CLUMPFIND algorithm \citep{Williams1994} with a method which simply selected cells above a given density, and found that the cloud mass spectra obtained with the latter method was more curved. Differences in slope, and the position of the peak, were also found for clump mass functions (CMFs) when using CLUMPFIND and dendrograms  \citep{Cheng2018}. These, and our work, suggest that the choice of clumpfinding algorithm will likely have some influence on the mass spectra obtained, highlighting the importance of using the same analysis technique to make comparisons.

We also compared other properties of the clouds, including velocity dispersion, size, rotation and virial relation using both the friends of friends algorithm and CPROPS. Again the properties are fairly similar in the observations and simulations. Particularly when comparing the clouds using the CPROPS algorithm, the observed GMCs lie almost intermediate between the clouds from the simulations with 10 and 20\% feedback. All the simulated and observed clouds found using CPROPS lie with virial parameters slightly above one, and on a similar virial relation, suggesting that the virial parameter may not be a particularly useful distinguishing parameter, but merely a consequence of the selection criteria for the clouds. The range of cloud angular momentum is also similar between all the simulations and observations, with the simulations having very comparable fractions of retrograde clouds to the observations. Using the pixel-by-pixel approach of \citet{Sun2018}, which avoids the need to discretise the gas into clouds, we again found good agreement between the properties found in the simulated and observed data. Our emission maps show some differences however, with the emission extending to larger radii in the simulated galaxies. This could be because we do not follow molecular chemistry in the simulations, and the gas at larger radii may be predominantly HI.

None of the simulated M33 galaxies provide a perfect match to the clouds properties of the actual M33. In some instances we obtain excellent agreement with one property (e.g. the cloud mass spectra of GASOLINE20 with CPROPS) but do not match so well other properties. The CPROPS comparisons suggest that the optimum level of feedback required to more precisely match the observation may be between that used in the simulations presented here i.e. around 15 \%. 

\section{Conclusions}
We have carried out a direct comparison between GMCs in an observed galaxy (M33) and simulated versions of the same galaxy. We found good agreement with the properties of the GMCs in the simulations, such as mass spectra, cloud rotations, velocity dispersions, virial relation, and those found in observations. This represents one of the first attempts to simulate cloud properties in a specific galaxy other than the Milky Way, and in particular the first such analysis for a spiral galaxy other than the Milky Way. Furthermore, we determined properties with the same method used as the observations, CPROPS, as well as a friends of friends algorithm. We also used a pixel-by-pixel approach, which is an alternative to decomposing the emission or density into clouds.

M33 is a smaller and more flocculent galaxy compared to the Milky Way. If we can reproduce the properties of molecular clouds in simulations of specific galaxies, that may be able to tell us what physics is important in reproducing the clouds, and ultimately star formation in those galaxies. We find strong agreement between the cloud properties in our simulations and the real M33. Our results, and those of \citet{Pettitt2018}, suggest that ultimately the main driver of molecular cloud properties is likely to be the gas surface density, and gravitational potential. \citet{Pettitt2018} found that there is no strong dependence of cloud properties on the mechanism of spiral arm formation, whether spiral arms are tidally induced or driven by underlying gravitational instabilities. In this paper we see that the variation in cloud masses with radius is dependent on the radial variation in gas surface density. The cloud mass spectrum, and its variation with surface density, radius, and level of star formation, appears to be a strong characteristic to test the simulations with observations. Other properties, such as cloud angular momentum and virial relation, also agree well between the simulations and observations, however these properties seem to be similar in other simulations and galaxies, suggesting they are not a distinguishing feature of a particular galaxy. 

We also examined the lifecycle of the clouds in our simulations. We see that the large majority of non star-forming clouds do go on to produce stars. This indicates that methods used by observers to estimate cloud lifetimes assuming that clouds spend a certain fraction of time without significant star formation, before having observable HII regions or stellar clusters, fits with the scenario we find in the simulations.

Our analysis also revealed differences between different clump-finding algorithms, in particular our simple friends of friends algorithm and CPROPS. The main difference was the shape of the mass spectra, which were much more curved with CPROPS. Such differences could be related to the spectral nature of CPROPS, and the tendency to group small clouds into larger objects. For much of the work we presented, we were mostly interested in the relative difference in GMC properties between different simulations, and the observations. However our results suggest that we should be cautious about over-interpreting the shape or slope of mass spectra when only one method has been applied to the data. 

\section*{Acknowledgments}
We thank the referee, Andreas Burkert, for helpful suggestions which improved the clarity of our paper. Calculations for this paper were performed on the ISCA High Performance Computing Service at the University of Exeter, and used the DiRAC Complexity system, operated by the University of Leicester IT Services, which forms part of the STFC DiRAC HPC Facility (www.dirac.ac.uk ). This equipment is funded by BIS National E-Infrastructure capital grant ST/K000373/1 and  STFC DiRAC Operations grant ST/K0003259/1. DiRAC is part of the National E-Infrastructure.

\appendix
\section{Further results for the FFGASOL10 population}
\begin{figure}
\centerline{\includegraphics[scale=0.4]{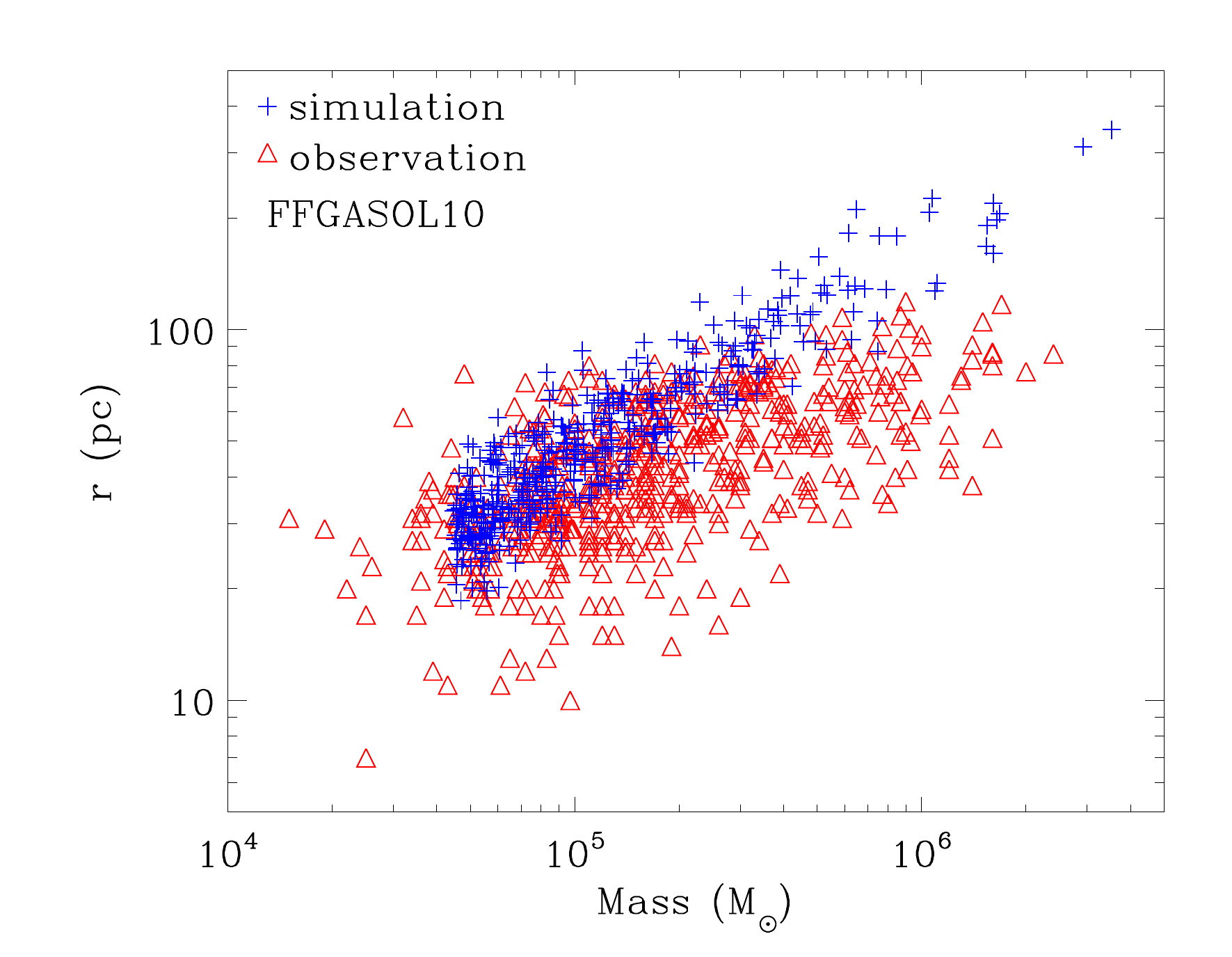}}
\centerline{\includegraphics[scale=0.4]{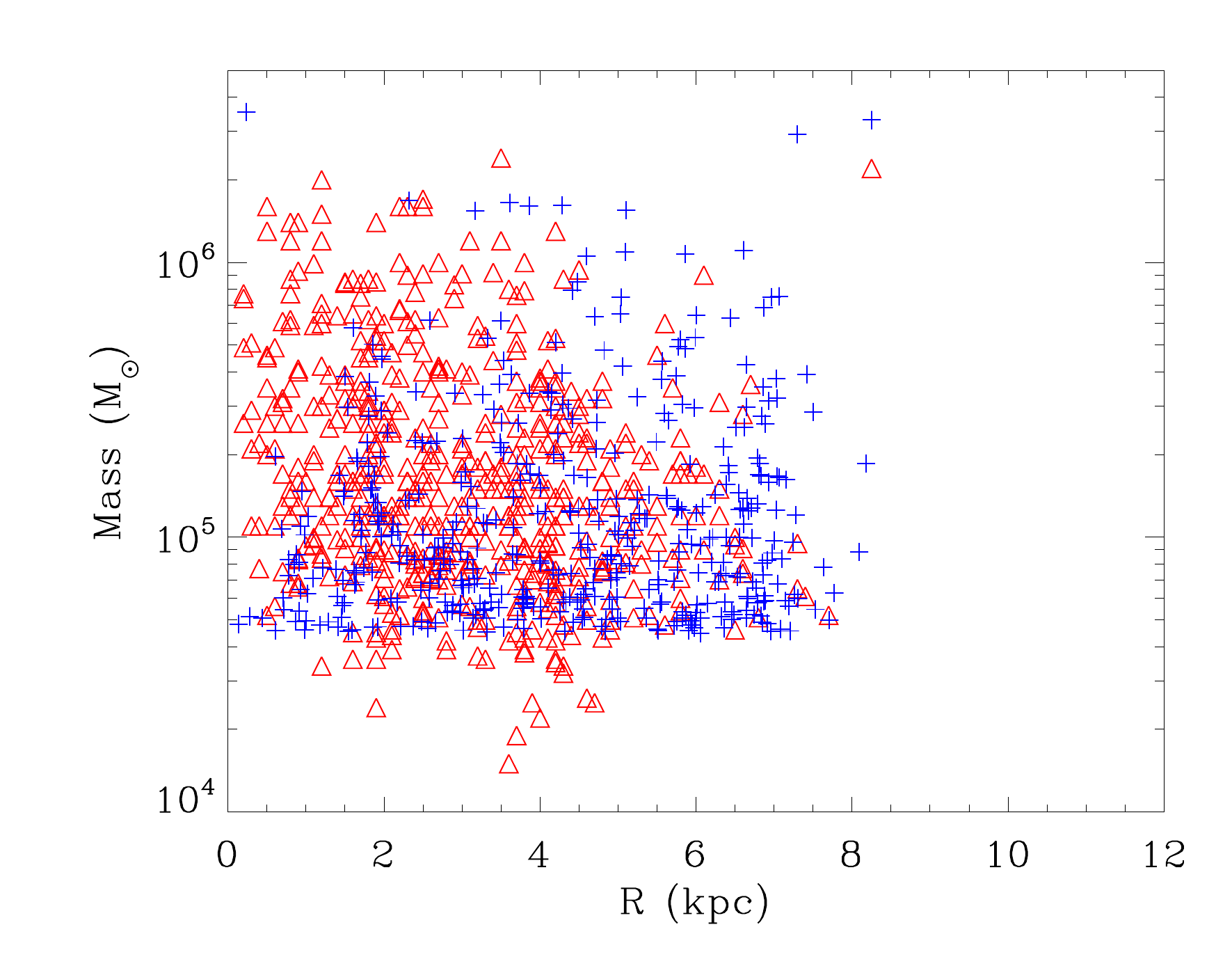}}
\centerline{\includegraphics[scale=0.4]{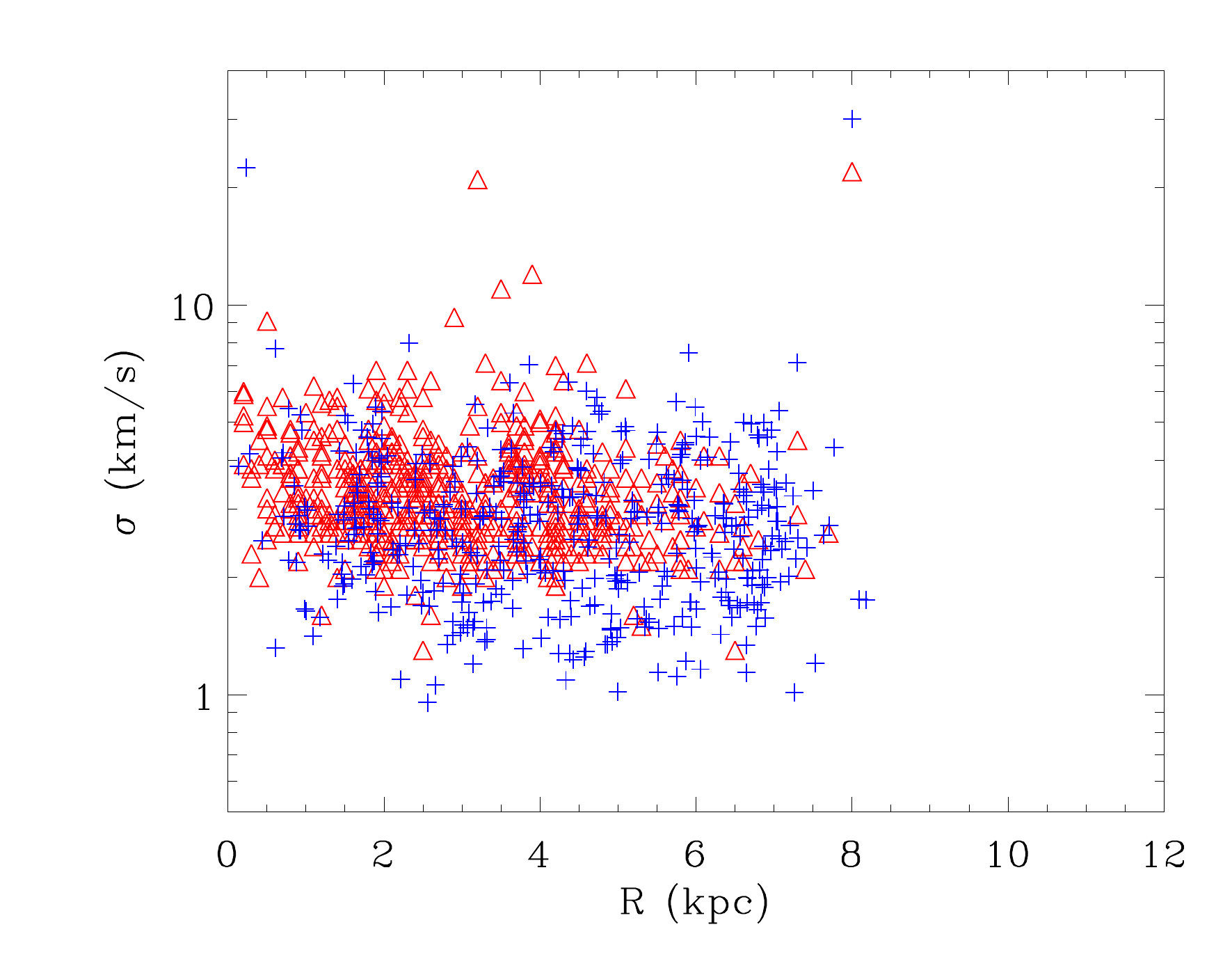}}
\caption{Cloud properties are shown for the FFGASOL10 clouds (radius and mass (top), mass versus galactic radius (centre) and velocity dispersion versus galactic radius (lower)). The observational results are also shown.}
\label{fig:ap1}
\end{figure}

\begin{figure}
\centerline{\includegraphics[scale=0.4]{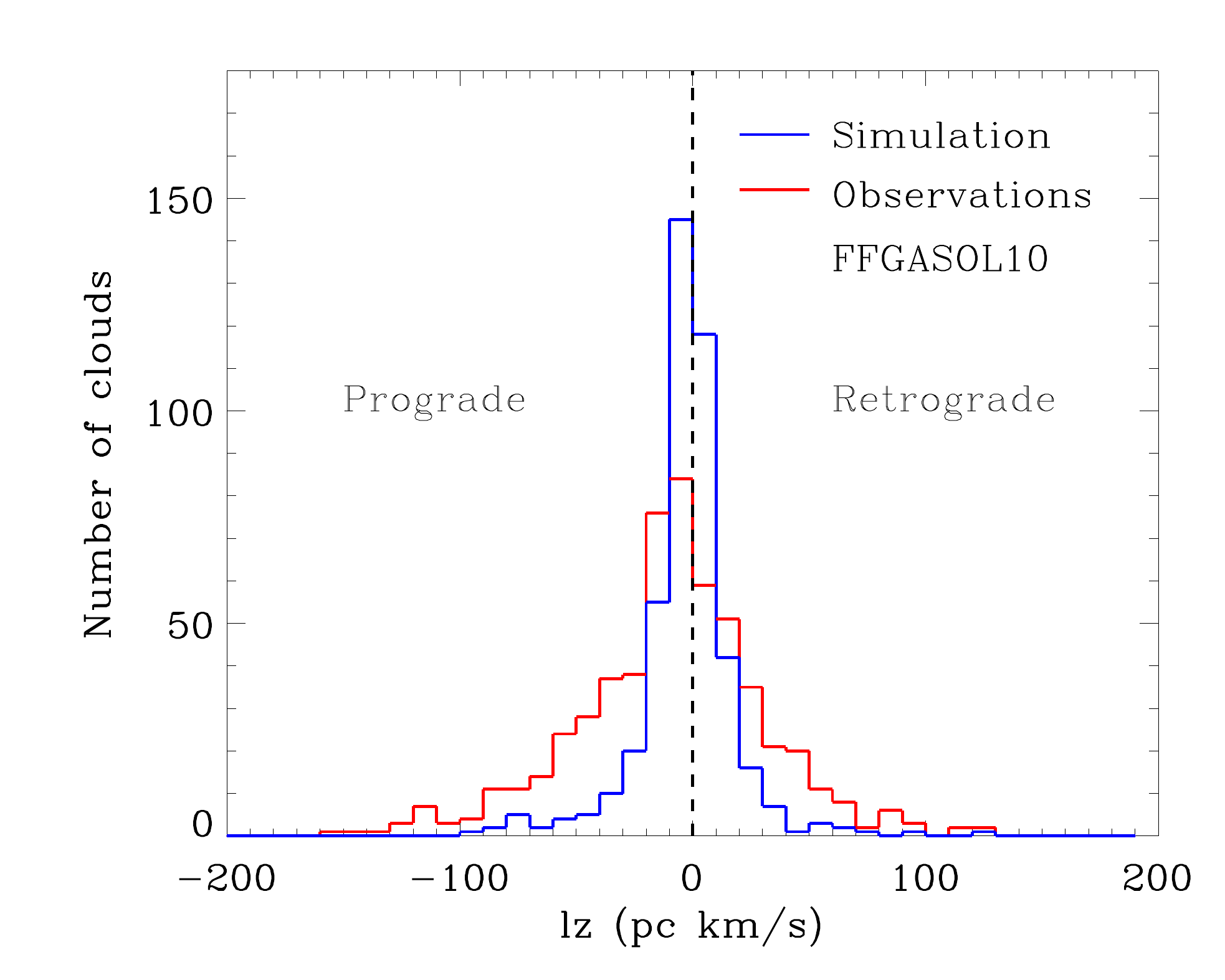}}
\caption{The angular momentum distribution of the clouds is shown for FFGASOL10 from the GASOLINE10 simulation. The distribution for the observations is also shown.}
\label{fig:ap2}
\end{figure}

In Figures~\ref{fig:ap1} and \ref{fig:ap2} we show further cloud properties for the FFGASOL10 clouds found using the friends of friends algorithm, which are discussed in the main text. These results are from the GASOLINE10 simulation.

\bibliographystyle{mn2e}
\bibliography{Dobbs}

\begin{thebibliography}{}

\bibitem[\protect\citeauthoryear{{Bate}, {Bonnell} \& {Price}}{{Bate}
  et~al.}{1995}]{Bate1995}
{Bate} M.~R.,  {Bonnell} I.~A.,    {Price} N.~M.,  1995, \mnras, 277, 362

\bibitem[\protect\citeauthoryear{{Bigiel}, {Leroy}, {Walter}, {Brinks}, {de
  Blok}, {Madore} \& {Thornley}}{{Bigiel} et~al.}{2008}]{Bigiel2008}
{Bigiel} F.,  {Leroy} A.,  {Walter} F.,  {Brinks} E.,  {de Blok} W.~J.~G.,
  {Madore} B.,    {Thornley} M.~D.,  2008, \aj, 136, 2846

\bibitem[\protect\citeauthoryear{{Braine}, {Rosolowsky}, {Gratier}, {Corbelli}
  \& {Schuster}}{{Braine} et~al.}{2018}]{Braine2018}
{Braine} J.,  {Rosolowsky} E.,  {Gratier} P.,  {Corbelli} E.,    {Schuster}
  K.-F.,  2018, \aap, 612, A51

\bibitem[\protect\citeauthoryear{{Cheng}, {Tan}, {Liu}, {Kong}, {Lim},
  {Andersen} \& {Da Rio}}{{Cheng} et~al.}{2018}]{Cheng2018}
{Cheng} Y.,  {Tan} J.~C.,  {Liu} M.,  {Kong} S.,  {Lim} W.,  {Andersen} M.,
  {Da Rio} N.,  2018, \apj, 853, 160

\bibitem[\protect\citeauthoryear{{Colombo}, {Hughes}, {Schinnerer}, {Meidt},
  {Leroy}, {Pety}, {Dobbs}, {Garc{\'{\i}}a-Burillo}, {Dumas}, {Thompson},
  {Schuster} \& {Kramer}}{{Colombo} et~al.}{2014}]{Colombo2014}
{Colombo} D.,  {Hughes} A.,  {Schinnerer} E.,  {Meidt} S.~E.,  {Leroy} A.~K.,
  {Pety} J.,  {Dobbs} C.~L.,  {Garc{\'{\i}}a-Burillo} S.,  {Dumas} G.,
  {Thompson} T.~A.,  {Schuster} K.~F.,    {Kramer} C.,  2014, \apj, 784, 3

\bibitem[\protect\citeauthoryear{{Colombo}, {Rosolowsky}, {Ginsburg},
  {Duarte-Cabral} \& {Hughes}}{{Colombo} et~al.}{2015}]{Colombo2015}
{Colombo} D.,  {Rosolowsky} E.,  {Ginsburg} A.,  {Duarte-Cabral} A.,
  {Hughes} A.,  2015, \mnras, 454, 2067

\bibitem[\protect\citeauthoryear{{Corbelli}, {Braine}, {Bandiera}, {Brouillet},
  {Combes}, {Druard}, {Gratier}, {Mata}, {Schuster}, {Xilouris} \&
  {Palla}}{{Corbelli} et~al.}{2017}]{Corbelli2017}
{Corbelli} E.,  {Braine} J.,  {Bandiera} R.,  {Brouillet} N.,  {Combes} F.,
  {Druard} C.,  {Gratier} P.,  {Mata} J.,  {Schuster} K.,  {Xilouris} M.,
  {Palla} F.,  2017, \aap, 601, A146

\bibitem[\protect\citeauthoryear{{Corbelli}, {Braine} \&
  {Giovanardi}}{{Corbelli} et~al.}{2019}]{Corbelli2019}
{Corbelli} E.,  {Braine} J.,    {Giovanardi} C.,  2019, arXiv e-prints

\bibitem[\protect\citeauthoryear{{Corbelli}, {Thilker}, {Zibetti}, {Giovanardi}
  \& {Salucci}}{{Corbelli} et~al.}{2014}]{Corbelli2014}
{Corbelli} E.,  {Thilker} D.,  {Zibetti} S.,  {Giovanardi} C.,    {Salucci} P.,
   2014, \aap, 572, A23

\bibitem[\protect\citeauthoryear{{de Vaucouleurs}, {de Vaucouleurs}, {Corwin}
  Jr., {Buta}, {Paturel} \& {Fouqu{\'e}}}{{de Vaucouleurs}
  et~al.}{1991}]{DeVauc1991}
{de Vaucouleurs} G.,  {de Vaucouleurs} A.,  {Corwin} Jr. H.~G.,  {Buta} R.~J.,
  {Paturel} G.,    {Fouqu{\'e}} P.,  1991

\bibitem[\protect\citeauthoryear{{Dobbs}}{{Dobbs}}{2008}]{Dobbs2008}
{Dobbs} C.~L.,  2008, \mnras, 391, 844

\bibitem[\protect\citeauthoryear{{Dobbs}, {Burkert} \& {Pringle}}{{Dobbs}
  et~al.}{2011}]{Dobbs2011new}
{Dobbs} C.~L.,  {Burkert} A.,    {Pringle} J.~E.,  2011, \mnras, 417, 1318

\bibitem[\protect\citeauthoryear{{Dobbs}, {Pettitt}, {Corbelli} \&
  {Pringle}}{{Dobbs} et~al.}{2018}]{Dobbs2018}
{Dobbs} C.~L.,  {Pettitt} A.~R.,  {Corbelli} E.,    {Pringle} J.~E.,  2018,
  \mnras

\bibitem[\protect\citeauthoryear{{Dobbs} \& {Pringle}}{{Dobbs} \&
  {Pringle}}{2013}]{Dobbs2013}
{Dobbs} C.~L.,  {Pringle} J.~E.,  2013, \mnras, 432, 653

\bibitem[\protect\citeauthoryear{{Dobbs}, {Pringle} \& {Duarte-Cabral}}{{Dobbs}
  et~al.}{2015}]{Dobbs2015}
{Dobbs} C.~L.,  {Pringle} J.~E.,    {Duarte-Cabral} A.,  2015, \mnras, 446,
  3608

\bibitem[\protect\citeauthoryear{{Druard}, {Braine}, {Schuster}, {Schneider},
  {Gratier}, {Bontemps}, {Boquien}, {Combes}, {Corbelli}, {Henkel}, {Herpin},
  {Kramer}, {van der Tak} \& {van der Werf}}{{Druard}
  et~al.}{2014}]{Druard2014}
{Druard} C.,  {Braine} J.,  {Schuster} K.~F.,  {Schneider} N.,  {Gratier} P.,
  {Bontemps} S.,  {Boquien} M.,  {Combes} F.,  {Corbelli} E.,  {Henkel} C.,
  {Herpin} F.,  {Kramer} C.,  {van der Tak} F.,    {van der Werf} P.,  2014,
  \aap, 567, A118

\bibitem[\protect\citeauthoryear{{Duarte-Cabral}, {Acreman}, {Dobbs},
  {Mottram}, {Gibson}, {Brunt} \& {Douglas}}{{Duarte-Cabral}
  et~al.}{2015}]{Duarte2015}
{Duarte-Cabral} A.,  {Acreman} D.~M.,  {Dobbs} C.~L.,  {Mottram} J.~C.,
  {Gibson} S.~J.,  {Brunt} C.~M.,    {Douglas} K.~A.,  2015, \mnras, 447, 2144

\bibitem[\protect\citeauthoryear{{Duarte-Cabral} \& {Dobbs}}{{Duarte-Cabral} \&
  {Dobbs}}{2016}]{Duarte2016}
{Duarte-Cabral} A.,  {Dobbs} C.~L.,  2016, \mnras, 458, 3667

\bibitem[\protect\citeauthoryear{{Duarte-Cabral} \& {Dobbs}}{{Duarte-Cabral} \&
  {Dobbs}}{2017}]{Duarte2017}
{Duarte-Cabral} A.,  {Dobbs} C.~L.,  2017, \mnras, 470, 4261

\bibitem[\protect\citeauthoryear{{Elmegreen}, {Elmegreen}, {Kaufman}, {Brinks},
  {Struck}, {Bournaud}, {Sheth} \& {Juneau}}{{Elmegreen}
  et~al.}{2017}]{Elmegreen2017}
{Elmegreen} D.~M.,  {Elmegreen} B.~G.,  {Kaufman} M.,  {Brinks} E.,  {Struck}
  C.,  {Bournaud} F.,  {Sheth} K.,    {Juneau} S.,  2017, \apj, 841, 43

\bibitem[\protect\citeauthoryear{{Engargiola}, {Plambeck}, {Rosolowsky} \&
  {Blitz}}{{Engargiola} et~al.}{2003}]{Engargiola2003}
{Engargiola} G.,  {Plambeck} R.~L.,  {Rosolowsky} E.,    {Blitz} L.,  2003,
  \apjs, 149, 343

\bibitem[\protect\citeauthoryear{{Faesi}, {Lada} \& {Forbrich}}{{Faesi}
  et~al.}{2018}]{Faesi2018}
{Faesi} C.~M.,  {Lada} C.~J.,    {Forbrich} J.,  2018, \apj, 857, 19

\bibitem[\protect\citeauthoryear{{Field} \& {Saslaw}}{{Field} \&
  {Saslaw}}{1965}]{Field1965}
{Field} G.~B.,  {Saslaw} W.~C.,  1965, \apj, 142, 568

\bibitem[\protect\citeauthoryear{{Freeman}, {Rosolowsky}, {Kruijssen},
  {Bastian} \& {Adamo}}{{Freeman} et~al.}{2017}]{Freeman2017}
{Freeman} P.,  {Rosolowsky} E.,  {Kruijssen} J.~M.~D.,  {Bastian} N.,
  {Adamo} A.,  2017, \mnras, 468, 1769

\bibitem[\protect\citeauthoryear{{Gardan}, {Braine}, {Schuster}, {Brouillet} \&
  {Sievers}}{{Gardan} et~al.}{2007}]{Gardan2007}
{Gardan} E.,  {Braine} J.,  {Schuster} K.~F.,  {Brouillet} N.,    {Sievers} A.,
   2007, \aap, 473, 91

\bibitem[\protect\citeauthoryear{{Gratier}, {Braine}, {Rodriguez-Fernandez},
  {Schuster}, {Kramer}, {Corbelli}, {Combes}, {Brouillet}, {van der Werf} \&
  {R{\"o}llig}}{{Gratier} et~al.}{2012}]{Gratier2012}
{Gratier} P.,  {Braine} J.,  {Rodriguez-Fernandez} N.~J.,  {Schuster} K.~F.,
  {Kramer} C.,  {Corbelli} E.,  {Combes} F.,  {Brouillet} N.,  {van der Werf}
  P.~P.,    {R{\"o}llig} M.,  2012, \aap, 542, A108

\bibitem[\protect\citeauthoryear{{Gratier et al.}}{{Gratier et
  al.}}{2010}]{Gratier2010}
{Gratier et al.} 2010, \aap, 522, A3

\bibitem[\protect\citeauthoryear{{Gratier et al.}}{{Gratier et
  al.}}{2017}]{Gratier2017}
{Gratier et al.} 2017, \aap, 600, A27

\bibitem[\protect\citeauthoryear{{Grisdale}, {Agertz}, {Renaud} \&
  {Romeo}}{{Grisdale} et~al.}{2018}]{Grisdale2018}
{Grisdale} K.,  {Agertz} O.,  {Renaud} F.,    {Romeo} A.~B.,  2018, \mnras,
  479, 3167

\bibitem[\protect\citeauthoryear{{Hughes}, {Meidt}, {Colombo}, {Schinnerer},
  {Pety}, {Leroy}, {Dobbs}, {Garc{\'{\i}}a-Burillo}, {Thompson}, {Dumas},
  {Schuster} \& {Kramer}}{{Hughes} et~al.}{2013}]{Hughes2013}
{Hughes} A.,  {Meidt} S.~E.,  {Colombo} D.,  {Schinnerer} E.,  {Pety} J.,
  {Leroy} A.~K.,  {Dobbs} C.~L.,  {Garc{\'{\i}}a-Burillo} S.,  {Thompson}
  T.~A.,  {Dumas} G.,  {Schuster} K.~F.,    {Kramer} C.,  2013, \apj, 779, 46

\bibitem[\protect\citeauthoryear{{Kaneko}, {Kuno} \& {Saitoh}}{{Kaneko}
  et~al.}{2018}]{Kaneko2018}
{Kaneko} H.,  {Kuno} N.,    {Saitoh} T.~R.,  2018, \apjl, 860, L14

\bibitem[\protect\citeauthoryear{{Kennicutt et al.}}{{Kennicutt et
  al.}}{2007}]{Kennicutt2007}
{Kennicutt et al.} 2007, \apj, 671, 333

\bibitem[\protect\citeauthoryear{{Khoperskov}, {Vasiliev}, {Ladeyschikov},
  {Sobolev} \& {Khoperskov}}{{Khoperskov} et~al.}{2016}]{Khoperskov2016}
{Khoperskov} S.~A.,  {Vasiliev} E.~O.,  {Ladeyschikov} D.~A.,  {Sobolev} A.~M.,
     {Khoperskov} A.~V.,  2016, \mnras, 455, 1782

\bibitem[\protect\citeauthoryear{{Koch}, {Rosolowsky}, {Lockman}, {Kepley},
  {Leroy}, {Schruba}, {Braine}, {Dalcanton}, {Johnson} \&
  {Stanimirovi{\'c}}}{{Koch} et~al.}{2018}]{Koch2018}
{Koch} E.~W.,  {Rosolowsky} E.~W.,  {Lockman} F.~J.,  {Kepley} A.~A.,  {Leroy}
  A.,  {Schruba} A.,  {Braine} J.,  {Dalcanton} J.,  {Johnson} M.~C.,
  {Stanimirovi{\'c}} S.,  2018, \mnras, 479, 2505

\bibitem[\protect\citeauthoryear{{Kreckel}, {Faesi}, {Kruijssen}, {Schruba},
  {Groves}, {Leroy}, {Bigiel}, {Blanc}, {Chevance}, {Herrera}, {Hughes},
  {McElroy}, {Pety}, {Querejeta}, {Rosolowsky}, {Schinnerer}, {Sun}, {Usero} \&
  {Utomo}}{{Kreckel} et~al.}{2018}]{Kreckel2018}
{Kreckel} K.,  {Faesi} C.,  {Kruijssen} J.~M.~D.,  {Schruba} A.,  {Groves} B.,
  {Leroy} A.~K.,  {Bigiel} F.,  {Blanc} G.~A.,  {Chevance} M.,  {Herrera} C.,
  {Hughes} A.,  {McElroy} R.,  {Pety} J.,  {Querejeta} M.,  {Rosolowsky} E.,
  {Schinnerer} E.,  {Sun} J.,  {Usero} A.,    {Utomo} D.,  2018, \apjl, 863,
  L21

\bibitem[\protect\citeauthoryear{{Larson}}{{Larson}}{1981}]{Larson1981}
{Larson} R.~B.,  1981, \mnras, 194, 809

\bibitem[\protect\citeauthoryear{{Leroy}, {Hughes}, {Schruba}, {Rosolowsky},
  {Blanc}, {Bolatto}, {Colombo}, {Escala}, {Kramer}, {Kruijssen}, {Meidt},
  {Pety}, {Querejeta}, {Sandstrom}, {Schinnerer}, {Sliwa} \& {Usero}}{{Leroy}
  et~al.}{2016}]{Leroy2016}
{Leroy} A.~K.,  {Hughes} A.,  {Schruba} A.,  {Rosolowsky} E.,  {Blanc} G.~A.,
  {Bolatto} A.~D.,  {Colombo} D.,  {Escala} A.,  {Kramer} C.,  {Kruijssen}
  J.~M.~D.,  {Meidt} S.,  {Pety} J.,  {Querejeta} M.,  {Sandstrom} K.,
  {Schinnerer} E.,  {Sliwa} K.,    {Usero} A.,  2016, \apj, 831, 16

\bibitem[\protect\citeauthoryear{{Mestel}}{{Mestel}}{1966}]{Mestel1966}
{Mestel} L.,  1966, \mnras, 131, 307

\bibitem[\protect\citeauthoryear{{Nguyen}, {Pettitt}, {Tasker} \&
  {Okamoto}}{{Nguyen} et~al.}{2018}]{Nguyen2018}
{Nguyen} N.~K.,  {Pettitt} A.~R.,  {Tasker} E.~J.,    {Okamoto} T.,  2018,
  \mnras, 475, 27

\bibitem[\protect\citeauthoryear{{Pan}, {Fujimoto}, {Tasker}, {Rosolowsky},
  {Colombo}, {Benincasa} \& {Wadsley}}{{Pan} et~al.}{2015}]{Pan2015}
{Pan} H.-A.,  {Fujimoto} Y.,  {Tasker} E.~J.,  {Rosolowsky} E.,  {Colombo} D.,
  {Benincasa} S.~M.,    {Wadsley} J.,  2015, \mnras, 453, 3082

\bibitem[\protect\citeauthoryear{{Pan}, {Fujimoto}, {Tasker}, {Rosolowsky},
  {Colombo}, {Benincasa} \& {Wadsley}}{{Pan} et~al.}{2016}]{Pan2016}
{Pan} H.-A.,  {Fujimoto} Y.,  {Tasker} E.~J.,  {Rosolowsky} E.,  {Colombo} D.,
  {Benincasa} S.~M.,    {Wadsley} J.,  2016, \mnras, 458, 2443

\bibitem[\protect\citeauthoryear{{Patel}, {Besla} \& {Sohn}}{{Patel}
  et~al.}{2017}]{Patel2017}
{Patel} E.,  {Besla} G.,    {Sohn} S.~T.,  2017, \mnras, 464, 3825

\bibitem[\protect\citeauthoryear{{Pettitt}, {Egusa}, {Dobbs}, {Tasker},
  {Fujimoto} \& {Habe}}{{Pettitt} et~al.}{2018}]{Pettitt2018}
{Pettitt} A.~R.,  {Egusa} F.,  {Dobbs} C.~L.,  {Tasker} E.~J.,  {Fujimoto} Y.,
    {Habe} A.,  2018, \mnras, 480, 3356

\bibitem[\protect\citeauthoryear{{Renaud}, {Bournaud} \& {Duc}}{{Renaud}
  et~al.}{2015}]{Renaud2015}
{Renaud} F.,  {Bournaud} F.,    {Duc} P.-A.,  2015, \mnras, 446, 2038

\bibitem[\protect\citeauthoryear{{Rosolowsky} \& {Leroy}}{{Rosolowsky} \&
  {Leroy}}{2006}]{Rosolowsky2006}
{Rosolowsky} E.,  {Leroy} A.,  2006, \pasp, 118, 590

\bibitem[\protect\citeauthoryear{{Schinnerer}, {Meidt}, {Pety}, {Hughes},
  {Colombo}, {Garc{\'{\i}}a-Burillo}, {Schuster}, {Dumas}, {Dobbs}, {Leroy},
  {Kramer}, {Thompson} \& {Regan}}{{Schinnerer} et~al.}{2013}]{Schinnerer2013}
{Schinnerer} E.,  {Meidt} S.~E.,  {Pety} J.,  {Hughes} A.,  {Colombo} D.,
  {Garc{\'{\i}}a-Burillo} S.,  {Schuster} K.~F.,  {Dumas} G.,  {Dobbs} C.~L.,
  {Leroy} A.~K.,  {Kramer} C.,  {Thompson} T.~A.,    {Regan} M.~W.,  2013,
  \apj, 779, 42

\bibitem[\protect\citeauthoryear{{Smith}, {Glover}, {Clark}, {Klessen} \&
  {Springel}}{{Smith} et~al.}{2014}]{Smith2014}
{Smith} R.~J.,  {Glover} S.~C.~O.,  {Clark} P.~C.,  {Klessen} R.~S.,
  {Springel} V.,  2014, \mnras, 441, 1628

\bibitem[\protect\citeauthoryear{{Stinson}, {Seth}, {Katz}, {Wadsley},
  {Governato} \& {Quinn}}{{Stinson} et~al.}{2006}]{Stinson2006}
{Stinson} G.,  {Seth} A.,  {Katz} N.,  {Wadsley} J.,  {Governato} F.,
  {Quinn} T.,  2006, \mnras, 373, 1074

\bibitem[\protect\citeauthoryear{{Sun}, {Leroy}, {Schruba}, {Rosolowsky},
  {Hughes}, {Kruijssen}, {Meidt}, {Schinnerer}, {Blanc}, {Bigiel}, {Bolatto},
  {Chevance}, {Groves}, {Herrera}, {Hygate}, {Pety}, {Querejeta}, {Usero} \&
  {Utomo}}{{Sun} et~al.}{2018}]{Sun2018}
{Sun} J.,  {Leroy} A.~K.,  {Schruba} A.,  {Rosolowsky} E.,  {Hughes} A.,
  {Kruijssen} J.~M.~D.,  {Meidt} S.,  {Schinnerer} E.,  {Blanc} G.~A.,
  {Bigiel} F.,  {Bolatto} A.~D.,  {Chevance} M.,  {Groves} B.,  {Herrera}
  C.~N.,  {Hygate} A.~P.~S.,  {Pety} J.,  {Querejeta} M.,  {Usero} A.,
  {Utomo} D.,  2018, \apj, 860, 172

\bibitem[\protect\citeauthoryear{{Tasker} \& {Tan}}{{Tasker} \&
  {Tan}}{2009}]{Tasker2009}
{Tasker} E.~J.,  {Tan} J.~C.,  2009, \apj, 700, 358

\bibitem[\protect\citeauthoryear{{Tosaki}, {Kohno}, {Harada}, {Tanaka},
  {Egusa}, {Izumi}, {Takano}, {Nakajima}, {Taniguchi} \& {Tamura}}{{Tosaki}
  et~al.}{2017}]{Tosaki2017}
{Tosaki} T.,  {Kohno} K.,  {Harada} N.,  {Tanaka} K.,  {Egusa} F.,  {Izumi} T.,
   {Takano} S.,  {Nakajima} T.,  {Taniguchi} A.,    {Tamura} Y.,  2017, \pasj,
  69, 18

\bibitem[\protect\citeauthoryear{{Utomo et al.}}{{Utomo et
  al.}}{2018}]{Utomo2018}
{Utomo et al.} 2018, \apjl, 861, L18

\bibitem[\protect\citeauthoryear{{van der Marel}, {Fardal}, {Sohn}, {Patel},
  {Besla}, {del Pino-Molina}, {Sahlmann} \& {Watkins}}{{van der Marel}
  et~al.}{2018}]{vanderMarel2018}
{van der Marel} R.~P.,  {Fardal} M.~A.,  {Sohn} S.~T.,  {Patel} E.,  {Besla}
  G.,  {del Pino-Molina} A.,  {Sahlmann} J.,    {Watkins} L.~L.,  2018, ArXiv
  e-prints

\bibitem[\protect\citeauthoryear{{Wadsley}, {Keller} \& {Quinn}}{{Wadsley}
  et~al.}{2017}]{Wadsley2017}
{Wadsley} J.~W.,  {Keller} B.~W.,    {Quinn} T.~R.,  2017, \mnras, 471, 2357

\bibitem[\protect\citeauthoryear{{Williams}, {de Geus} \& {Blitz}}{{Williams}
  et~al.}{1994}]{Williams1994}
{Williams} J.~P.,  {de Geus} E.~J.,    {Blitz} L.,  1994, \apj, 428, 693

\end{thebibliography}
\bsp
\label{lastpage}
\end{document}